\documentclass[aps,prc,twocolumn,amsmath,amssymb,nobibnotes,superscriptaddress,longbibliography]{revtex4-2}
\usepackage{textcomp}
\usepackage{gensymb}
\usepackage{bm}
\usepackage{graphicx}
\usepackage{dcolumn}
\usepackage{array}[=2016-10-06]
\usepackage{hyperref}
\hypersetup{
    colorlinks=true,
    linkcolor=blue,
    filecolor=blue,
    urlcolor=blue,
    citecolor= blue,}
\usepackage{bm}
\newcommand{\beqn}{\vspace{0.5em}\begin{eqnarray}}
\newcommand{\eeqn}{\end{eqnarray}\par\vspace{0.5em}\noindent}

\begin{document}
\title{Low-lying states of neutron-rich $N=50$ isotones in multi-reference covariant density functional theory}
 \author{X. Y. Wu}
\email[]{xywu@jxnu.edu.cn}
 \author{M. L. Mai}
 \author{D. Q. Zhu}
 \author{W. H. Liu}
 \author{Z. M. Liu}
 \affiliation{School of Physics, Jiangxi Normal University, Nanchang 330022, China}
 \author{J. Xiang}
 \email[]{jxiang@cqnu.edu.cn}
 \affiliation{College of Physics and Electronic Engineering, Chongqing Normal University, Chongqing 401331, China}


\begin{abstract}
Predictions for the systematics of low-lying states in neutron-rich $N=50$ isotones from $^{80}\mathrm{Zn}$ to $^{70}\mathrm{Ca}$, where experimental information remains scarce, exhibit substantial model dependence. We investigate these nuclei using multi-reference covariant density functional theory based on a relativistic energy density functional, in which nuclear wave functions are constructed as superpositions of symmetry-restored mean-field states with different quadrupole deformations. The low-lying states of $^{78}\mathrm{Ni}$ are reasonably reproduced, and a prolate rotational band built on the second $0^+$ state is predicted in addition to a near-spherical ground state. As the proton number decreases toward $Z=20$, the low-energy structure exhibits a nonmonotonic evolution from spherical to strongly deformed and subsequently back to spherical shapes, consistent with valence-space shell-model predictions. In particular, the emergence of deformed ground states in $^{76}\mathrm{Fe}$ and $^{74}\mathrm{Cr}$ is accompanied by enhanced quadrupole collectivity and stronger shape mixing. The results provide a coherent microscopic description of the structural evolution from the doubly magic $^{78}\mathrm{Ni}$ to the neutron-rich $N=50$ isotones.

\end{abstract}

\maketitle
%
 \section{INTRODUCTION}
 \label{Introduction}
Low-lying states provide a sensitive probe of nuclear shell structure. The magic numbers that define closed proton or neutron shells have been a cornerstone of nuclear physics~\cite{Erler12,Wienholtz13,Ye25}. However, the evolution of shell structure away from stability provides stringent tests of modern many-body theories~\cite{Bender03,Vretenar05,Otsuka20} and crucial inputs for $r$-process nucleosynthesis~\cite{Pfeiffer01,Arnould07,Baruah2008,Cowan21}, as manifested in the erosion of magic numbers~\cite{Bastin07,Ding26}, the emergence of islands of inversion~\cite{Gray23,Rocchini23,Zhou25}, and the appearance of shape coexistence~\cite{Wood92,Heyde11}. The $N=50$ isotones, ranging from the doubly magic $^{78}\mathrm{Ni}$ ($Z=28$) to more neutron-rich nuclei, offer an ideal testing ground for these phenomena, challenging both experimental and theoretical studies. Recent experiments at next-generation radioactive-ion-beam facilities have investigated the neutron-rich $N=50$ shell closure~\cite{Orlandi15,Welker17,Vajta18} and established low-lying states in $^{79}\mathrm{Cu}$~\cite{Olivier17} and $^{79}\mathrm{Zn}$~\cite{Nies23}. Taniuchi \textit{et al.}~\cite{Taniuchi2019} reported the first spectroscopy of $^{78}\mathrm{Ni}$, revealing $E(2^+_1)=2.60$~MeV and a tentative $2^+_2$ at $2.91$~MeV, which indicate spherical closure and deformation, respectively, and suggest shape coexistence. Evidence for shape coexistence has also been identified in measurements in neighboring nuclei~\cite{Gottardo16,Yang16}. Although enormous experimental efforts have been made, experimental data in this region remain scarce. Consequently, the evolution of shell structure in low-lying states along the neutron-rich $N=50$ isotonic chain remains an open question~\cite{Nowacki21,Taniuchi26}.

 Theoretically, various advanced approaches have been applied to study the low-lying states of neutron-rich $N=50$ isotones. Valence-space-based approaches, such as large-scale shell model (LSSM)~\cite{Nowacki2016}, Monte Carlo shell model (MCSM)~\cite{Tsunoda2014,Kumar26}, realistic shell model (RSM)~\cite{Li23}, and valence-space density matrix renormalization group (VS-DMRG)~\cite{Tichai24}, along with full-space-based methods including coupled-cluster (CC)~\cite{Hagen2016,Hu2024} and in-medium similarity renormalization group (IM-SRG)~\cite{Taniuchi2019,Simonis17}, have reproduced the shell structure and shape coexistence in $^{78}\mathrm{Ni}$ and explored the structural evolution along the neutron-rich $N=50$ isotonic chain from $^{80}\mathrm{Zn}$ to $^{70}\mathrm{Ca}$. In valence-space approaches, model-space truncation influences low-lying spectra, while effective charges in transition operators affect electric transition strengths, particularly in neutron-rich nuclei. As a result, quantitative predictions for electric transition strengths, particularly $B(E2)$ and $E0$ strengths, exhibit substantial model dependence. Different choices of valence space lead to discrepancies in the calculated $B(E2)$ strengths across the neutron-rich $N=50$ isotonic chain.

 Within the energy density functional (EDF) theoretical framework, the beyond-mean-field (BMF) approach based on mean-field calculations in the full single-particle space, namely the five-dimensional collective Hamiltonian (5DCH)~\cite{Delaroche10}, has been employed using the Gogny D1S interaction to perform systematic calculations of low-lying states. In contrast to the methods discussed above, the 5DCH approach underestimates the $E(2^+_1)$ value and fails to capture the $N=50$ shell evolution predicted by the valence-space shell-model from $^{80}$Zn toward $^{70}$Ca. Moreover, among other EDF-based BMF approaches, the quasiparticle random-phase approximation (QRPA)~\cite{Peru2008} overestimates the $2^+_1$ energy of $^{78}\mathrm{Ni}$; the cranking approximation Bohr Hamiltonian (CABH)~\cite{Girod88} yields a reasonable $2^+_1$ energy but overestimates the $4^+_1$ energy of $^{78}\mathrm{Ni}$ and has not been applied along the neutron-rich $N=50$ isotonic chain. None of these three EDF-based BMF approaches, however, reproduces the deformed rotational band and shape coexistence in $^{78}\mathrm{Ni}$. In addition, the $E0$ transition strengths associated with shape coexistence have not yet been investigated from $^{80}$Zn toward $^{70}$Ca. These issues provide the motivation for the present work.

 In this paper, we adopt the multi-reference covariant density functional theory (MR-CDFT), i.e., the generator coordinate method (GCM) with particle-number and angular-momentum projections~\cite{Yao09,Yao10,Wu15,Yao20,Yao22}, to investigate the low-lying states in this region. This BMF approach restores the broken symmetries and accounts for quantum shape fluctuations of the mean-field reference states within the full single-particle space and without adjustable parameters, thereby avoiding the need for effective charges in calculating electric transition strengths. Our calculations reproduce key spectroscopic fingerprints of the doubly magic closure in $^{78}\mathrm{Ni}$ and establish the coexistence of a spherical ground-state band and a deformed rotational band. The successful description of $^{78}\mathrm{Ni}$ encourages us to extend the study to the low-lying states and related electric transition strengths of $N=50$ isotones from $^{80}\mathrm{Zn}$ to $^{70}\mathrm{Ca}$. The low-energy structure exhibits a nonmonotonic evolution from spherical to strongly deformed and back to spherical shapes, in line with valence-space shell-model predictions but in contrast to the 5DCH results. Our results provide robust constraints on collectivity and deformation in this key region, offering a unified and self-consistent BMF description of structural evolution along the neutron-rich $N=50$ isotonic chain.

 This paper is organized as follows: Sec.~\ref{Method} briefly outlines the theoretical method. Numerical details are given in Sec.~\ref{Numerical}. Results and discussions are presented in Sec.~\ref{Results}, and the conclusions are summarized in Sec.~\ref{conclusion}.
%

%
\section{THE THEORETICAL METHODS}
 \label{Method}
 In the MR-CDFT framework, the nuclear collective wave function is constructed via configuration mixing of particle-number and angular-momentum projected mean-field states
  \beqn
  \label{wavefun}
  \vert \Psi^{JNZ}_k \rangle
  =\sum_{q} f^{J}_k (q)\hat P^J_{MK} \hat P^N\hat P^Z\vert \Phi(q)\rangle  \, ,
  \eeqn
 where $k$ labels different collective states with the same  angular momentum $J$. The operators $\hat P^J_{MK}$, $\hat P^{N}$, and $\hat P^{Z}$ project onto angular momentum $J$, neutron $N$, and proton $Z$ numbers, respectively. The reference states $\vert\Phi(q)\rangle$ denotes a set of relativistic mean-field plus Bardeen-Cooper-Schrieffer (RMF+BCS) states with deformation parameters $q=(\beta,\gamma)$~\cite{Gambhir90,Ring96,Meng06}. In the case of axisymmetric quadrupole deformation with $\gamma=0$, the deformation parameter $q$ reduces to $\beta$.  
 Applying the variational principle to the weight functions $f^{J}_k (\beta)$ yields the Hill-Wheeler-Griffin (HWG) equation~\cite{Hill53,Griffin57,Ring80}
  \beqn
  \label{HWGE}
  \sum_{\beta^\prime} \left[ \mathcal{H}^{J}(\beta,\beta^\prime)
          - E_k ^{J} \, \mathcal{N}^{J}(\beta,\beta^\prime)
           \right] f_k ^{J}(\beta^\prime) = 0.
   \eeqn
 Here $\mathcal{O}^{J}(\beta,\beta^\prime)=\langle\Phi(\beta)\vert\hat O\hat P^J_{MK}\hat P^N\hat P^Z\vert \Phi(\beta^\prime)\rangle$, which corresponds to the norm kernel for $\hat O=1$ and to the energy kernel for $\hat O=\hat H$. Solving the HWG equation yields the collective weight functions and the energy spectrum, from which electric multipole transition strengths are calculated.

 The electric quadrupole transition strength from the initial state
 $(J_i,k_i)$ to the final state $(J_f,k_f)$ is calculated by
 \begin{eqnarray}
  &&B(E2; J_i,k_i\rightarrow J_f,k_f)\nonumber\\
    &=& \frac{1}{2J_i+1}
    \left|\sum_{\beta^\prime,\,\beta}f_{k_f}^{J_{f}\,\ast}(\beta^\prime) \, \langle J_f,\beta^\prime\vert\vert \hat Q_{2}\vert\vert J_i,\beta\rangle
        f_{k_i}^{J_{i}}(\beta)\,\right|^2,\nonumber\\
 \label{BE2}
 \end{eqnarray}
 with the reduced transition matrix element
 \begin{eqnarray}
   &&\langle J_f,\beta^\prime\vert\vert \hat Q_{2}\vert\vert J_i,\beta\rangle\nonumber\\
   &=&\frac{(2J_f+1)(2J_i+1)}{2}
   \sum_{M=-2}^{+2}
   \left(
   \begin{array}{ccc}
     J_f  &  2     &J_i \\
     0    & M      &-M \\
   \end{array}
    \right) \nonumber\\
  &&  \times\int_0^\pi \sin\theta\, d\theta\, d_{-M0}^{J_i\ast}(\theta)\langle\Phi(\beta^\prime) \vert\hat Q_{2M}e^{i\theta\hat J_y}\hat P^N \hat P^Z\vert\Phi(\beta)\rangle \, ,\nonumber\\
  \label{Q2}
 \end{eqnarray}
 where $\hat Q_{2M}\equiv er^2Y_{2M}$ is the electric quadrupole moment operator.
 In addition, the electric monopole transition from the $0^+_2$ to the $0^+_1$ state is given by
 \beqn%
 \label{E0}
 \rho^2(E0; 0^+_2 \rightarrow 0^+_1) = \left| \frac{ \langle 0^+_2 | \hat{T}(E0) | 0^+_1 \rangle }{e R_0^2}\right|^2,
 \eeqn%
 with $R_0 = r_0 A^{1/3}$ and the transition matrix element
 \begin{eqnarray}\label{TE0}
 &&\langle 0_2^+ | \hat{T}(E0) | 0_1^+ \rangle \nonumber\\
 &=& \frac{1}{2}\,
 \sum_{\beta^\prime,\,\beta} f_{0_2^+}^{J_2=0\,\ast}(\beta^\prime) f_{0_1^+}^{J_1=0}(\beta) \nonumber\\
 &&\times\int_0^\pi \sin\theta \, d\theta\, \langle \Phi(\beta^\prime) | \hat{T}(E0) e^{i\theta \hat J_y} \hat{P}^{N}\hat{P}^{Z}| \Phi(\beta) \rangle\, ,
 \end{eqnarray}
 where $\hat{T}(E0) = \sum_{i} e_i r_i^2$. Since the full single-particle space is taken into account, no extra effective charges are needed to calculate the electric transition strengths; instead, the bare nucleon charges $e_p=1$ and $e_n=0$ are directly adopted (see Refs.~\cite{Yao09,Yao10,Wu15,Yao20,Yao22}).
%

\section{NUMERICAL DETAILS}
 \label{Numerical}
 In the constrained triaxial RMF calculations, parity, x-simplex symmetry, and time-reversal invariance are imposed. The Dirac equation is solved by expanding the Dirac spinors, separately for large and small components, in the basis of eigenfunctions of a three-dimensional harmonic oscillator in Cartesian coordinates.
 Ten major shells are found to be sufficient for the nuclei under consideration~\cite{Yao14}. We use the energy density functional (EDF) derived from the relativistic point-coupling Lagrangian PC-PK1~\cite{Zhao10}, which has been used to study nuclear ground states~\cite{Zhang22} and low-lying states~\cite{Wu15,Quan17,Wu24}. Pairing correlations between the nucleons are treated within the BCS approximation using a density-independent $\delta$-force with a smooth pairing window with a smooth cutoff factor~\cite{Krieger90}. The strength parameters of the pairing force are $V_n = -349.5$ and $V_p = -330$ MeV$\cdot\text{fm}^3$ for neutrons and protons, respectively. In the kernel calculations, the Gauss-Legendre quadrature is used for the integral over Euler angle $\theta$. The number of mesh points for the Euler angle $\theta$ and gauge angle $\varphi_{\tau=n,p}$ in the interval $[0, \pi]$ is chosen as 12 and 7 for the angular-momentum and particle-number projections, respectively. The Pfaffian method is used to determine the phase of the norm overlaps~\cite{Robledo09}.

%
%
\section{RESULTS and DISCUSSIONS}
 \label{Results}
 \begin{figure}[htbp]
    \centering
    \includegraphics[scale = 0.48]{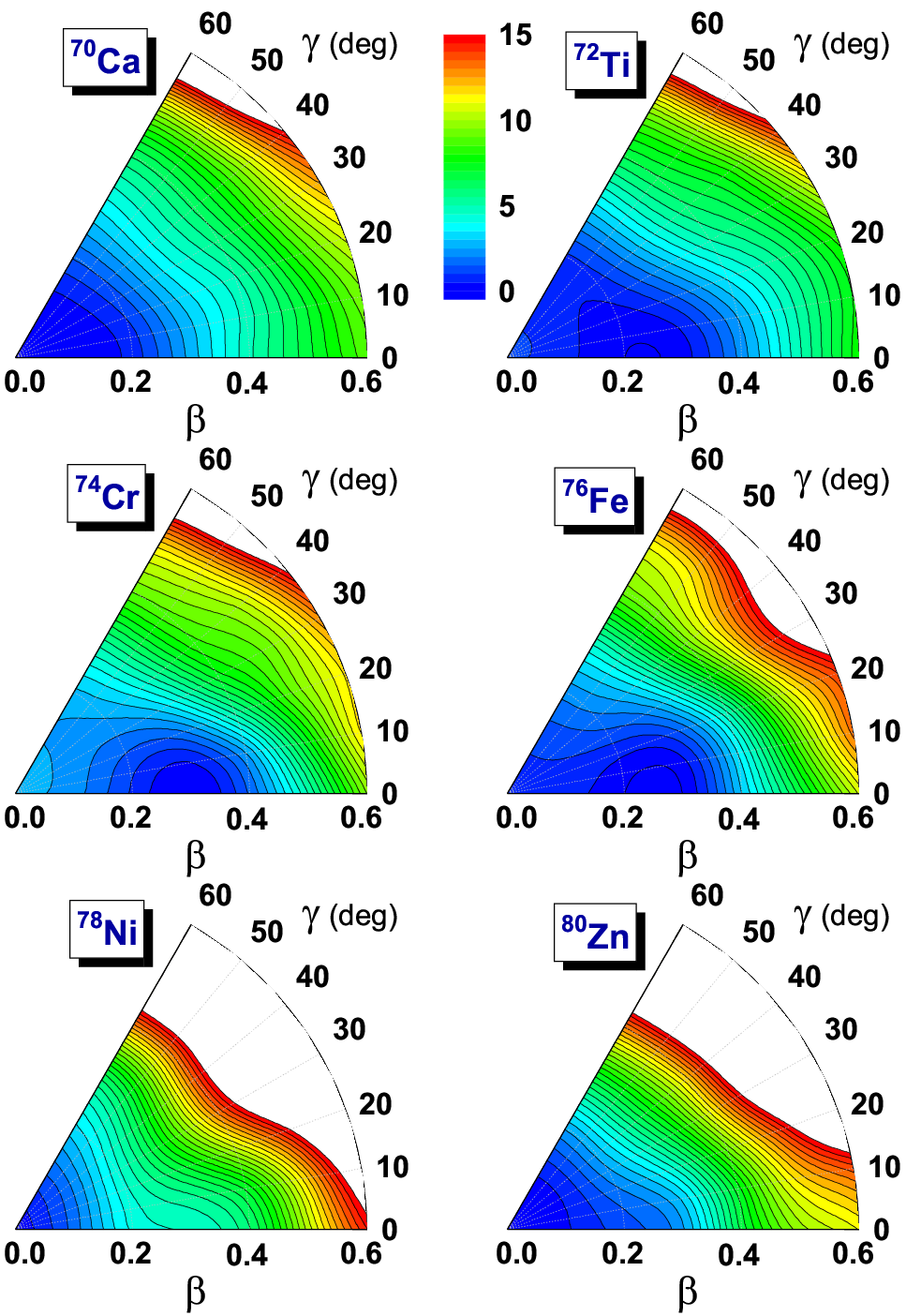}
     \caption{Potential energy surfaces in the $(\beta,\gamma)$ plane for neutron-rich $N=50$ isotones from RMF+BCS calculations using the EDF PC-PK1. Energies are normalized to the absolute minimum; contour interval is 0.5 MeV.}
    \label{fig1}
 \end{figure}
 Fig.~\ref{fig1} illustrates the evolution of shell structure along the neutron-rich $N=50$ isotonic chain via potential energy surfaces (PESs) in the $(\beta,\gamma)$ plane, obtained from RMF+BCS calculations with the EDF PC-PK1. The PESs reveal a gradual transition from a soft spherical minimum in $^{70}\mathrm{Ca}$ to a deformed prolate shape in $^{74}\mathrm{Cr}$, abruptly interrupted by the doubly magic $^{78}\mathrm{Ni}$ with a rigid spherical minimum. For $N=50$ isotones with $Z\ge28$, the deformed relativistic Hartree-Bogoliubov theory in continuum with PC-PK1 predicts spherical minima~\cite{Zhang22}. In contrast, Gogny D1S~\cite{Delaroche10} predicts spherical minima for all $N=50$ isotones from $^{72}\mathrm{Ti}$ to $^{80}\mathrm{Zn}$. Notably, removing two protons from $^{78}\mathrm{Ni}$ to $^{76}\mathrm{Fe}$ induces weak prolate softness, while adding two protons to $^{80}\mathrm{Zn}$ preserves sphericity, underscoring the robust $N=50$ shell closure along the Zn chain, in agreement with Ref.~\cite{Shiga16}.

 The RMF+BCS calculations, i.e., at the mean-field level of the CDFT, reveal the shell evolution as reflected in the ground-state properties. To account for quantum shape fluctuations and symmetry restoration, we employ the MR-CDFT approach to investigate the structural evolution of low-lying states along the neutron-rich $N=50$ isotonic chain at the beyond-mean-field level.
 Since there are no triaxial minima in the PESs obtained with the EDF PC-PK1 (see Fig.~\ref{fig1}) and with the Gogny EDF D1S~\cite{Delaroche10}, and given that three-dimensional projected GCM calculations on top of the CDFT are computationally demanding, all subsequent BMF calculations are restricted to axial quadrupole deformation.

 \begin{figure}[htbp]
    \centering
    \includegraphics[scale = 0.44]{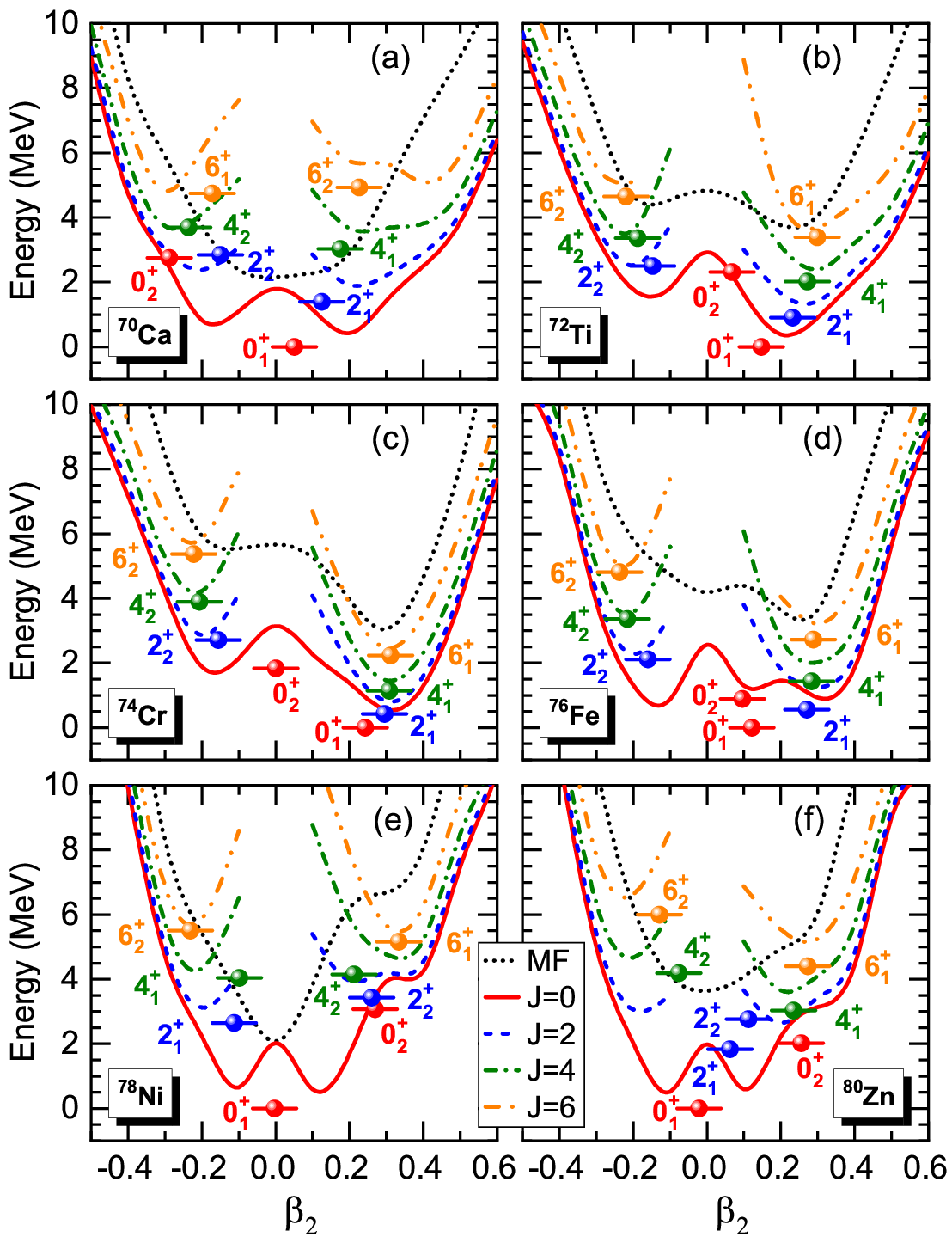}
    \caption{Total energy (normalized to the $0^+_1$ state) of the $N=50$ isotones as a function of intrinsic quadrupole deformation for mean-field (MF) and particle-number and angular-momentum projected states with $J=0, 2, 4, 6$. The solid bullets and horizontal bars indicate the lowest GCM solutions, shown at their average deformation.}
    \label{fig2}
 \end{figure}

 Fig.~\ref{fig2} displays the projected PECs for neutron-rich $N=50$ isotones. For the spherical nuclei $^{70}\mathrm{Ca}$, $^{78}\mathrm{Ni}$, and $^{80}\mathrm{Zn}$, the $J=0$ PEC exhibits two near-degenerate minima at prolate and oblate shapes close to sphericity. Projection generally deepens the RMF minima and extends the deformed region, as seen in $^{76}\mathrm{Fe}$ [Fig.~\ref{fig2}(d)], lying between spherical $^{78}\mathrm{Ni}$ and deformed $^{74}\mathrm{Cr}$. The absence of energy gain at $\beta=0$ for $^{78}\mathrm{Ni}$ after projection [Fig.~\ref{fig2}(e)] reflects pairing collapse due to the $Z=28$ and $N=50$ shell closures. Furthermore, solving the HWG equation yields discrete GCM states $J_\alpha^\pi$ with their average deformations $\overline{\beta}_2=\sum_\beta \beta|g_\alpha^J(\beta)|^2$~\cite{Ring80}. These states form two low-lying bands built on $0_1^+$ and $0_2^+$. In $^{80}\mathrm{Zn}$, the two rotational bands exhibit a shape crossing with increasing $J$ [Fig.~\ref{fig2}(f)]. In $^{70}\mathrm{Ca}$ and $^{72}\mathrm{Ti}$, near-degenerate $0_2^+$ and $2_2^+$ states form a $\beta$-vibrational band, reflecting the softness of the corresponding PESs [Figs.~\ref{fig2}(a,b)]. In contrast, the prolate ground-state band in $^{74}\mathrm{Cr}$ [Fig.~\ref{fig2}(c)] exhibits evident rotational character, which disappears in $^{78}\mathrm{Ni}$ [Fig.~\ref{fig2}(e)], where a spherical shell closure is predicted. Additionally, a low-lying rotational band built on the deformed $0^+$ state near the prolate minimum of the projected $J=0$ PEC emerges in $^{78}\mathrm{Ni}$, indicating shape coexistence.

\begin{figure}[htbp]
    \centering
    \includegraphics[scale = 0.45]{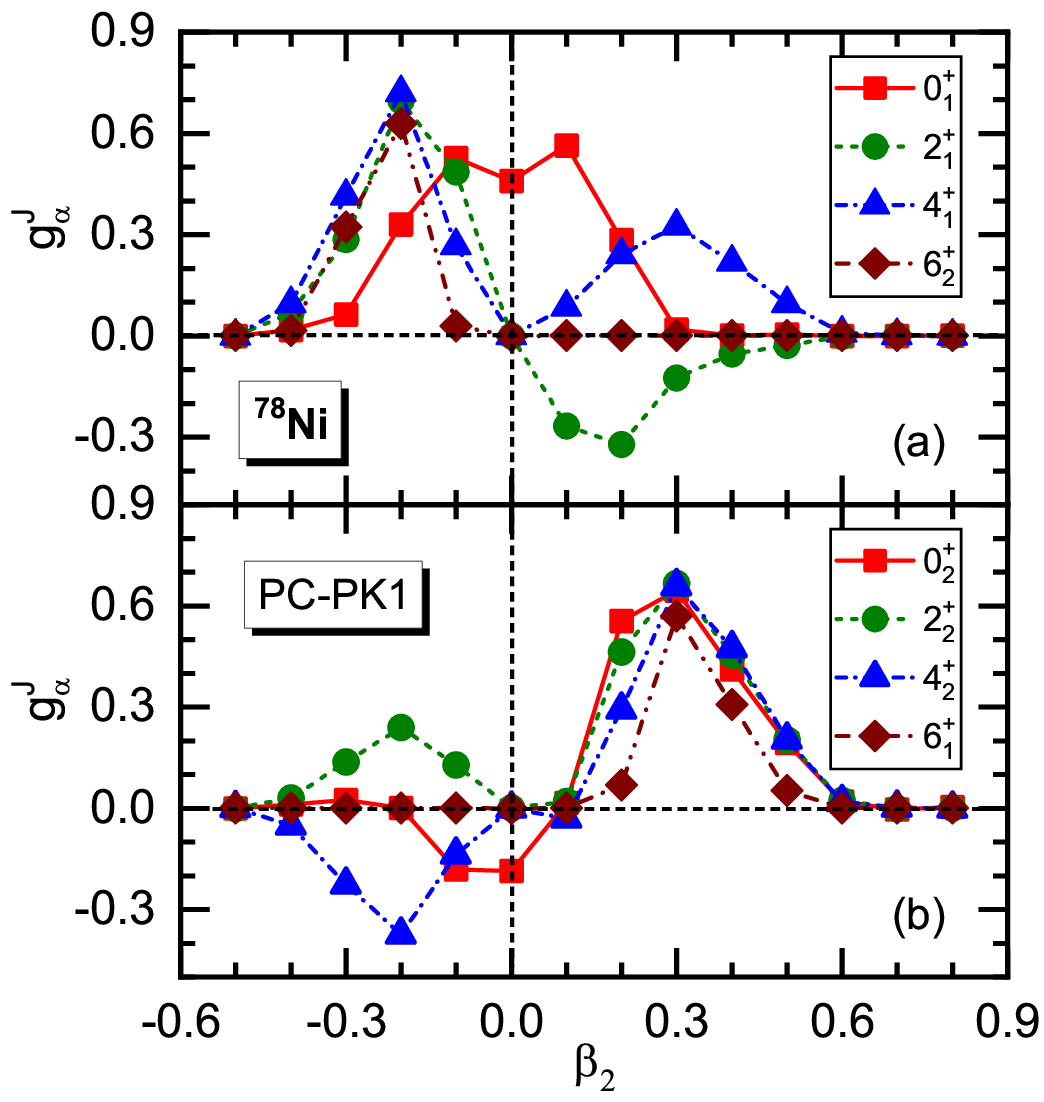}
     \caption{Collective wave functions of $^{78}\mathrm{Ni}$ as a function of the quadrupole deformation parameter $\beta_2$, calculated using the relativistic energy density functional PC-PK1. Panel (a) shows the wave functions for the $0^+$, $2^+$, $4^+$, and $6^+$ states of the spherical band; panel (b) shows those for the deformed band.}
    \label{fig3}
\end{figure}
 To elucidate the components of the excited states in the spherical and deformed bands obtained from our GCM calculations, we further present the wave-function distributions. Fig.~\ref{fig3} displays the collective wave functions of $^{78}\mathrm{Ni}$ as a function of the quadrupole deformation parameter $\beta_2$.
 The $0^+_1$ collective wave function in Fig.~\ref{fig3}(a) peaks around $\beta_2\simeq0$, indicating a spherical ground state. The $4^+_1$ wave function is similar to that of the $2^+_1$ state, both exhibiting peaks at $\beta_2\simeq\pm0.20$, forming a slightly deformed anharmonic vibrational band near the spherical shape from deformation perturbations.
 For the $6^+_2$ state built on the $4^+_1$ state, the wave function is dominated by the oblate component. Conversely, as shown in Fig.~\ref{fig3}(b), the wave functions of the $0^+_2$, $2^+_2$, $4^+_2$, and $6^+_1$ states are centered around $\beta_2\simeq0.30$, establishing a low-lying rotational band in $^{78}\mathrm{Ni}$. As found in LSSM and MCSM calculations, this prolate deformed band originates from the intruder $\nu 0g_{7/2}$ orbital across the $N=50$ shell gap~\cite{Taniuchi2019}, as illustrated in the single-particle level diagram of $^{78}\mathrm{Ni}$~\cite{Nowacki21}.

 \begin{figure*}[htbp]
    \centering
    \includegraphics[scale = 0.32]{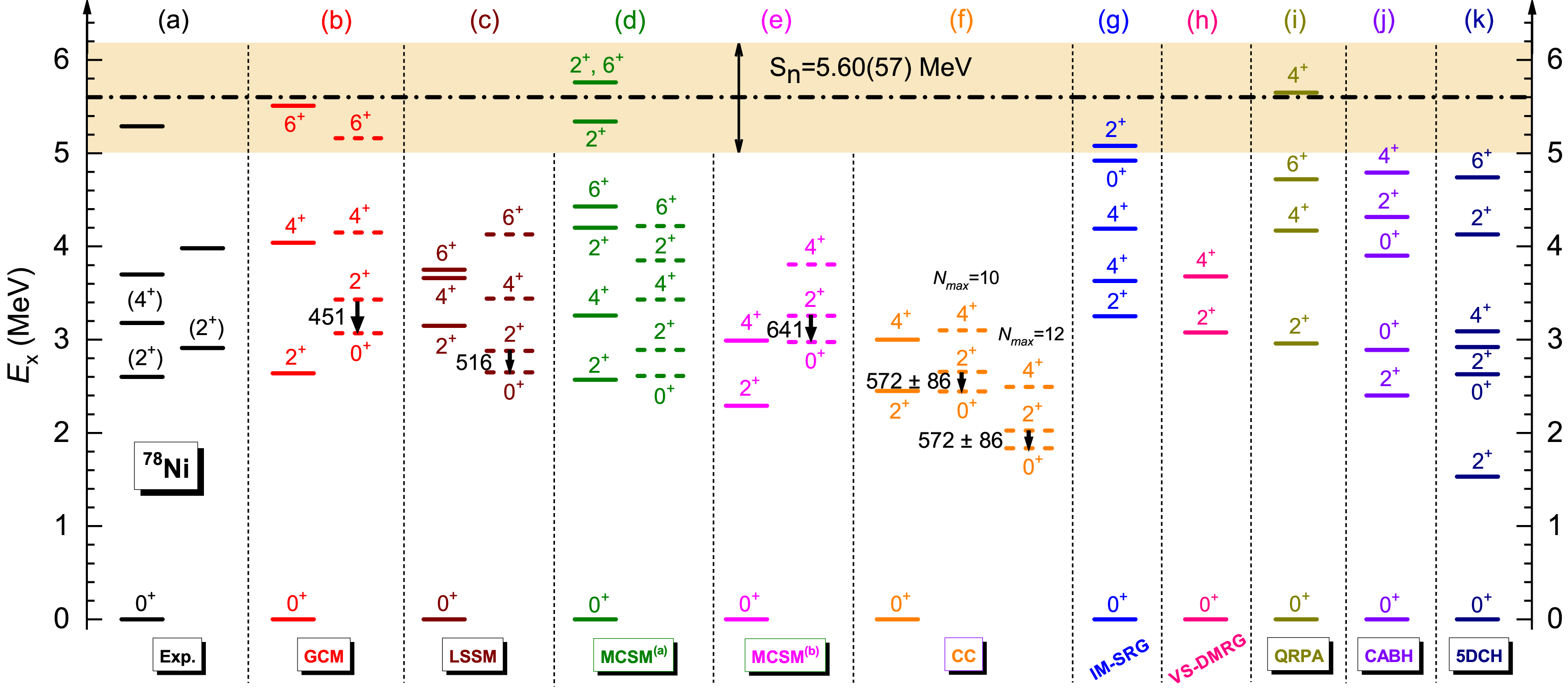}
    \caption{Comparison of theoretical predictions with the experimental spectrum of $^{78}\mathrm{Ni}$. (a) Experimental data from Ref.~\cite{Taniuchi2019}. (b) Our GCM results with the EDF PC-PK1. Other theoretical results are from: (c) LSSM~\cite{Nowacki2016}, (d) MCSM$^{\text{(a)}}$~\cite{Tsunoda2014}, (e) MCSM$^{\text{(b)}}$~\cite{Kumar26}, (f) CC~\cite{Hagen2016,Hu2024}, (g) IM-SRG~\cite{Simonis17,Taniuchi2019}, (h) VS-DMRG~\cite{Tichai24}, (i) QRPA (quasiparticle random-phase approximation)~\cite{Peru2008}, (j) CABH (cranking approximation Bohr Hamiltonian)~\cite{Girod88}, and (k) 5DCH~\cite{Delaroche10}. Deformed-band states predicted by GCM, LSSM, MCSM, and CC are indicated by dashed lines. The black dash-dotted line and light-yellow band mark the evaluated neutron separation energy $S_n = 5.60(57)$~MeV and its $1\sigma$ uncertainty~\cite{wang2021ame}. Arrows denote the deformed‑band $B(E2;2^+_2\to0^+_2)$ values (in $e^2\text{fm}^4$).}
    \label{fig4}
 \end{figure*}

 Fig.~\ref{fig4} displays the low-lying spectrum of $^{78}\mathrm{Ni}$. Our GCM calculations are consistent with LSSM, MCSM, and CC, revealing two distinct bands and supporting the interpretation that the spherical band is dominated by valence-space configurations, while the deformed band originates from particle-hole (\textit{ph}) excitations across the $Z=28$ and $N=50$ shell closures. The GCM result for $E(2_1^+)$, 2.64~MeV, agrees well with the experimental value (2.60), compared with LSSM (3.15)~\cite{Nowacki2016}, MCSM$^{\text{(a)}}$ (2.57)~\cite{Tsunoda2014}, MCSM$^{\text{(b)}}$ (2.30)~\cite{Kumar26}, CC (2.45)~\cite{Hagen2016,Hu2024}, IM-SRG (3.35)~\cite{Simonis17,Taniuchi2019}, VS-DMRG (3.08)~\cite{Tichai24}, QRPA (2.96)~\cite{Peru2008}, CABH (2.40)~\cite{Girod88}, and 5DCH (1.53)~\cite{Delaroche10}. The 5DCH result is notably lower, indicating a quenched shell closure. Nevertheless, all models except 5DCH yield $E(2_1^+) \ge 2.30$~MeV, significantly higher than the neighboring $^{76}\mathrm{Ni}$ (0.99~MeV) and $^{80}\mathrm{Zn}$ (1.49~MeV)~\cite{NNDC3}, confirming the persistent $N=50$ shell closure at $^{78}\mathrm{Ni}$. The detailed excitation energies of the $0^+$, $2^+$, and $4^+$ states are summarized in Table~\ref{tab1}. Moreover, we notice that two $J^\pi = 6^+$ excited states at $E = 5.60~\text{MeV}$ and $5.16~\text{MeV}$ are predicted by our GCM calculations. These values are within the neutron separation energy, $S_n = 5.60(57)~\text{MeV}$ from the evaluated nuclear mass table~\cite{wang2021ame} and align closely with the tentative experimental state at $E = 5.29~\text{MeV}$~\cite{Taniuchi2019}.
\begin{table*}[htbp]
    \centering
\caption{\label{tab1}
Low-spin $0^+, 2^+$, and $4^+$ excited state energies of $^{78}\mathrm{Ni}$ from experimental data and theoretical models, extracted from Fig.~\ref{fig4}. All energy values are in MeV.}
    \begin{ruledtabular}
    \begin{tabular}{cccccccccccc}
State&Experiment&GCM&LSSM&MCSM$^{\text{(a)}}$ &MCSM$^{\text{(b)}}$ &CC($N_\text{max}=10$) &IM-SRG &VS-DMRG &QRPA&CABH&5DCH\\ \hline
0$^+$&  0         &0    &0     &0    &0    &0    &0    &0    &0    &0    &0       \\
0$^+$&            &3.07 &2.65  &2.61 &2.98 &2.44 &4.92 &     &     &2.89 &2.63    \\
2$^+$&2.60        &2.64 &3.15  &2.57 &2.29 &2.45 &3.25 &3.08 &2.96 &2.40 &1.53    \\
2$^+$&2.91        &3.43 &2.88  &2.89 &3.26 &2.66 &5.08 &     &     &4.32 &2.92    \\
4$^+$&(3.18)      &4.04 &3.66  &3.26 &2.99 &3.00 &3.63 &3.68 &4.17 &4.79 &3.09    \\
4$^+$&(3.98)      &4.15 &3.44  &3.43 &3.81 &3.10 &4.19 &     &5.65 &     &        \\
\end{tabular}
\end{ruledtabular}
\end{table*}
\begin{table*}[htbp]
 \caption{\label{tab2}
 Low-lying properties of the deformed bands for the $N=50$ isotones, compared with LSSM calculations~\cite{Nowacki2016}. Excitation energies $E_x$ are in MeV, $B(E2)$ strengths in $e^2\text{fm}^4$, and spectroscopic quadrupole moments $Q_s$ in $e\text{fm}^2$.}
 \begin{ruledtabular}
 \begin{tabular}{cccccccccccccccccccc}
   &{Isotones}&{Method}& \multicolumn{3}{c}{$E_x$ (MeV) } & \multicolumn{3}{c}{$B(E2)$ ($e^2$fm$^4$)} &\multicolumn{3}{c}{$Q_s$($e$fm$^2$)}    \\
  \cline{4-6\  \ }  \cline{7-9\  \ } \cline{10-12\  \ }
 &  &  &\multicolumn{1}{c}{$2^+$}   & \multicolumn{1}{c}{$4^+$} & \multicolumn{1}{c}{$6^+$} &  \multicolumn{1}{c}{$2^{+}\to0^{+}$}   &\multicolumn{1}{c}{$4^{+}\to2^{+}$}   & \multicolumn{1}{c}{$6^{+}\to4^{+}$} &\multicolumn{1}{c}{$2^+$}   & \multicolumn{1}{c}{$4^+$} & \multicolumn{1}{c}{$6^+$} \\
 \hline
  &$^{78}$Ni   &LSSM    & 2.88    & 3.44   & 4.14   & 516   & 783   & 1021  &-39   & -65  & -75   \\
  &            &GCM     & 3.43   & 4.15    & 5.16   & 451   & 763   & 823   &-41   & -44  & -74   \\

  &$^{76}$Fe   &LSSM    & 0.43    & 1.05   & 1.90   & 314   & 707   & 780   &-45   & -57  & -63   \\
  &            &GCM     & 0.55   & 1.43    & 2.73   &  251  & 534   & 604   &-38   & -49  & -54   \\

  &$^{74}$Cr   &LSSM    & 0.24    & 0.72   & 1.38   & 630   & 911   & 1004  &-51   & -66  & -74   \\
  &            &GCM     & 0.42   & 1.14    & 2.23   & 324   & 507   & 570   &-37   & -48  & -53   \\

  &$^{72}$Ti   &LSSM    & 0.41    & 1.02   & 1.78   & 321   & 506   & 580   &-34   & -45  & -53   \\
  &            &GCM     & 0.90   & 2.02    & 3.39   & 147   & 271   & 348   &-23   & -33  & -39   \\

  &$^{70}$Ca   &LSSM    & 0.91    & 1.80   & 2.56   & 119   & 194   & 5     & -3   & +8   & +8    \\
  &            &GCM     & 1.39    & 3.04   & 4.94   & 33    & 73    & 73    &-7    & -14  & -24   \\
%
 \end{tabular}
 \end{ruledtabular}
 \end{table*}

 For the second $0^+$ state, IM-SRG predicts a considerably higher excitation energy than the other model calculations, likely due to missing collective correlations. Moreover, the intraband $B(E2;2^+_2\to0^+_2)=451~e^2\text{fm}^4$ predicted from GCM is consistent with LSSM ($516$)~\cite{Nowacki2016}, MCSM ($641$)~\cite{Kumar26}, and CC ($572\pm86$)~\cite{Hu2024} values, whereas the interband $B(E2;2^+_2\to0^+_1)$ strength is negligible ($0.01~e^2\text{fm}^4$) in our model.  In addition, for the deformed band of $^{78}\mathrm{Ni}$, our GCM predicts
 $R_{42}=E(4^+_2)/E(2^+_2)=3.01$, close to the rigid-rotor limit of $10/3$ and in line with LSSM (3.43)~\cite{Nowacki2016}, MCSM$^{\text{(a)}}$ (2.93)~\cite{Tsunoda2014}, MCSM$^{\text{(b)}}$ (2.95)~\cite{Kumar26}, and CC (3.10 for $N_\text{max}=10$, 3.50 for $N_\text{max}=12$)~\cite{Hu2024}, supporting its rotational character.

 Furthermore, additional $B(E2)$ values for the deformed bands of the $N=50$ isotones are summarized in Table~\ref{tab2}, along with the excitation energies $E_x$ and spectroscopic quadrupole moments $Q_s$, and
 are compared with LSSM calculations~\cite{Nowacki2016}. The large $B(E2)$ strengths and negative $Q_s$ values indicate that these states in $^{78}$Ni form a prolate rotational band. Our GCM predictions are in qualitative agreement with the LSSM results, except for a notable deviation in $^{70}$Ca. Although both approaches predict a negative $Q_s(2^+)$ for $^{70}$Ca, the $Q_s(4^+)$ and $Q_s(6^+)$ values they yield differ in magnitude by factors of $2\sim3$ and have opposite signs. Note that the predictions for $^{70}$Ca should be regarded as rough estimates. A reliable description of extremely neutron-rich calcium isotopes remains challenging because such a description requires realistic nuclear interactions, many-body correlations, and coupling to the scattering continuum. Relativistic and nonrelativistic EDF calculations predict the calcium drip line to be at the neutron-rich isotopes $^{72}$Ca and $^{70}$Ca, respectively~\cite{Meng02,Neufcourt19}. Moreover, realistic Gamow shell-model calculations including continuum coupling suggest a possible shell closure at $^{70}$Ca~\cite{Li20}.

 \begin{figure}[htbp]
    \centering
    \includegraphics[scale = 0.34]{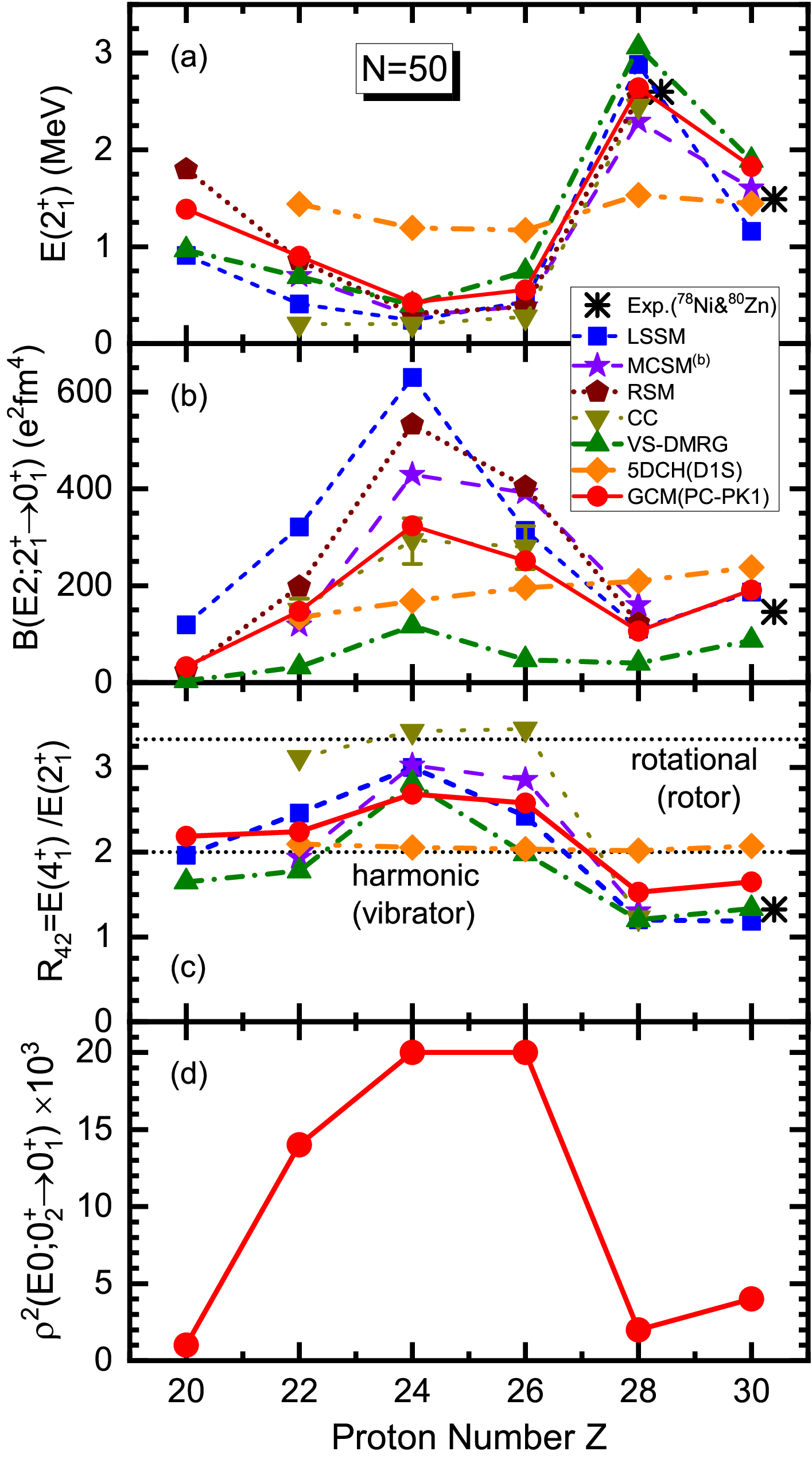}
    \caption{Panel (a): Excitation energies $E(2^+_1)$ (MeV). Panel (b): Electric quadrupole transition strengths $B(E2;2^+_1\to0^+_1)$ ($e^2\text{fm}^4$).
    Panel (c): Energy ratios $R_{42}=E(4^+_1)/E(2^+_1)$.
    Panel (d): Electric monopole transition strengths $\rho^2(E0;0^+_2\to0^+_1)$ for the $N=50$ isotones. Results are shown from LSSM (blue squares)~\cite{Nowacki2016}, MCSM$^{\text{(b)}}$ (violet stars)~\cite{Kumar26}, RSM (wine pentagons)~\cite{Li23}, CC (dark yellow down-triangles)~\cite{Hagen2016,Hu2024}, VS-DMRG (olive up-triangles)~\cite{Tichai24}, 5DCH (orange diamonds)~\cite{Delaroche10}, and GCM (red circles, this work). In panel (c), horizontal dotted lines at 2.00 and 3.33 mark the vibrational (harmonic oscillator) and rotational (rigid rotor) limits, respectively. Experimental data for $^{78}\mathrm{Ni}$ and $^{80}\mathrm{Zn}$ (black asterisks) are taken from Refs.~\cite{Taniuchi2019,Shiga16,Cortes18}.}
    \label{fig5}
 \end{figure}
 Fig.~\ref{fig5} shows the collective excitation systematics of the ground-state band for neutron-rich $N=50$ isotones from $^{80}\mathrm{Zn}$ to $^{70}\mathrm{Ca}$: (a) excitation energies $E(2^+_1)$, (b) electric quadrupole transition strengths $B(E2;2^+_1\to0^+_1)$, (c) energy ratios $R_{42}=E(4^+_1)/E(2^+_1)$, and (d) electric monopole transition strengths $\rho^2(E0;0^+_2\to0^+_1)$. Experimental data for $^{80}\mathrm{Zn}$ are from Refs.~\cite{Shiga16,Cortes18}. As shown in Fig.~\ref{fig5}(a), our GCM yields the highest $E(2^+_1)$ value for $^{78}\mathrm{Ni}$ among the $N=50$  isotonic chain, supporting the magicity of $^{78}\mathrm{Ni}$. In contrast, the low $E(2^+_1)$ values for $^{76}\mathrm{Fe}$ and $^{74}\mathrm{Cr}$ (around 0.50~MeV) indicate enhanced collectivity. These trends are well exhibited by LSSM~\cite{Nowacki2016}, MCSM$^{\text{(b)}}$~\cite{Kumar26}, RSM~\cite{Li23}, CC~\cite{Hagen2016,Hu2024}, and VS-DMRG~\cite{Tichai24}, whereas 5DCH~\cite{Delaroche10} approach fails to reproduce the $N=50$ shell closure at $^{78}\mathrm{Ni}$ and does not exhibit the structural evolution from $^{80}\mathrm{Zn}$ to $^{72}\mathrm{Ti}$.

 The $B(E2;2^+_1\to0^+_1)$ values in Fig.~\ref{fig5}(b) follow the expected inverse correlation with $E(2^+_1)$. Our GCM result for $^{80}\mathrm{Zn}$ ($191~e^2\text{fm}^4$) agrees with LSSM ($187~e^2\text{fm}^4$) and is slightly larger than the experimental value ($146~e^2\text{fm}^4$)~\cite{Shiga16,Cortes18}. For $^{78}\mathrm{Ni}$, our GCM gives $B(E2;2^+_1\to0^+_1)=106~e^2\text{fm}^4$, consistent with LSSM (110)~\cite{Nowacki2016}, MCSM$^{\text{(b)}}$ (160)~\cite{Kumar26}, and RSM (122)~\cite{Li23}. This value is close to the experimental benchmark for the spherical $^{56}\mathrm{Ni}$ ($99\pm24~e^2\text{fm}^4$)~\cite{Yurkewicz04}, supporting a spherical ground state, while the 5DCH result (209~$e^2\text{fm}^4$)~\cite{Delaroche10} is larger than the other model values. The $B(E2)$ value peaks at $^{74}\mathrm{Cr}$, suggesting collective rotational excitations and shell erosion~\cite{Nowacki2016,Li23}. All models except 5DCH yield qualitatively similar trends, but with large quantitative discrepancies around $^{74}\mathrm{Cr}$. The substantial model dependence of the $B(E2;2^+_1\to0^+_1)$ strengths highlights the importance of choosing a proper valence space for $B(E2)$ calculations. Our GCM calculations are consistent with the full-space CC model. In contrast, although the full single-particle space is taken into account, the 5DCH exhibits a monotonic trend, failing to capture the rapid onset of collective effects in the mid-shell nucleus $^{74}\mathrm{Cr}$, which lies between the non-collective isotones $^{70}\mathrm{Ca}$ and $^{78}\mathrm{Ni}$.

 The $R_{42}=E(4^+_1)/E(2^+_1)$ ratios in Fig.~\ref{fig5}(c) trace the evolution of collectivity along the $N=50$ isotonic chain. Our GCM results agree well with shell-model calculations in both trend and magnitude. Except for 5DCH, all models yield $R_{42}$ well below 2.0 only for $^{78}\mathrm{Ni}$ and $^{80}\mathrm{Zn}$, and their high $E(2^+_1)$ and low $B(E2)$ values together support a persistent $N=50$ shell closure far from stability. In contrast, GCM yields $ R_{42}\simeq2.69 $ for $^{74}\mathrm{Cr}$, close to the VS-DMRG value of 2.81 and qualitatively consistent with LSSM (3.00)~\cite{Nowacki2016}, MCSM$^{\text{(b)}}$ (3.03)~\cite{Tsunoda2014}, and CC (3.43)~\cite{Hu2024}, indicating a transition toward rotational collectivity. 

 The $2^+_1$ excitation energy, $B(E2;2^+_1\to0^+_1)$ strength, and $R_{42}$ ratio together indicate that the $N=50$ shell closure persists in $^{78}\mathrm{Ni}$ but breaks down in $^{76}\mathrm{Fe}$ and $^{74}\mathrm{Cr}$, where deformation and collectivity set in. This pattern mirrors the $N=40$ island of inversion, where the $N=40$ shell closure persists in $^{68}\mathrm{Ni}$~\cite{Sorlin02} but disappears in $^{66}\mathrm{Fe}$ and $^{64}\mathrm{Cr}$~\cite{Aoi09,Rother11,Baugher12}, suggesting that $^{76}\mathrm{Fe}$ and $^{74}\mathrm{Cr}$ may mark the onset of a similar island of inversion at $N=50$~\cite{Li23}.
 \begin{figure}[htbp]
    \centering
    \includegraphics[scale = 0.38]{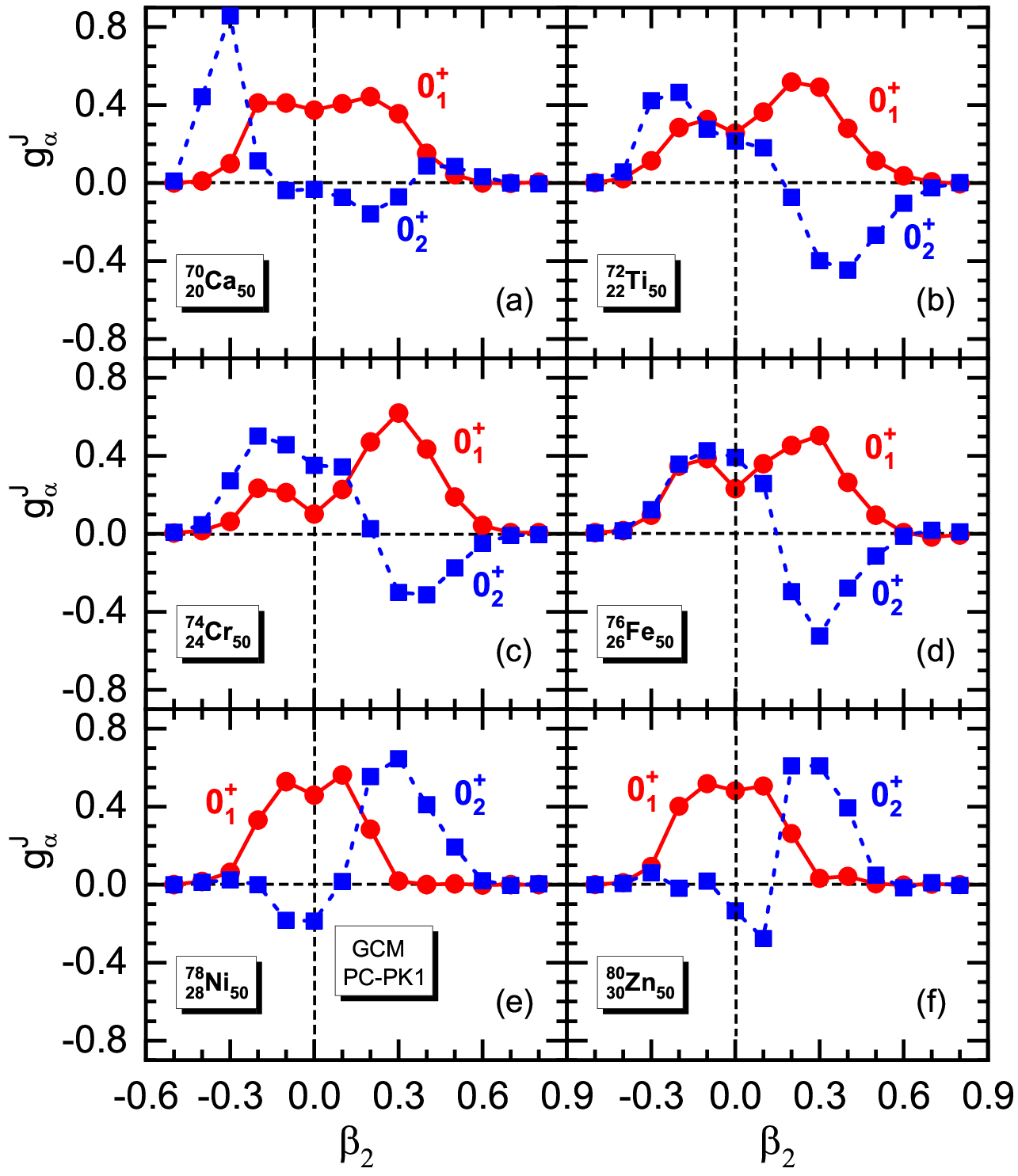}
 \caption{Collective wave functions of $N=50$ isotones as a function of quadrupole deformation $\beta_2$, obtained from GCM calculations with the relativistic EDF PC-PK1.}
   \label{fig6}
 \end{figure}

 Electric monopole transitions in Fig.~\ref{fig5}(d) are sensitive probes of shape coexistence along the $N=50$ isotonic chain. For $^{78}\mathrm{Ni}$, despite the distinct charge radii of the two $0^+$ states, our calculations yield a weak $\rho^2(E0;0^+_2\to0^+_1)\simeq2\times10^{-3}$ transition strength. In contrast, enhanced collectivity in $^{74}\mathrm{Cr}$ and $^{76}\mathrm{Fe}$ yields much larger $\rho^2(E0)$ values of about $20\times10^{-3}$. The rapid evolution of the $E0$ transition strength along the neutron-rich $N=50$ isotonic chain is understood from the $\beta_2$ dependence of the collective wave functions, shown in Fig.~\ref{fig6}. Although the root-mean-square charge radii differ by $\Delta\langle r_{\mathrm{ch}}^2\rangle^{1/2}=0.06~\mathrm{fm}$ between the spherical $0_1^+$ and prolate deformed $0_2^+$ states in $^{78}\mathrm{Ni}$, the small $\rho^2(E0)$ value indicates that wave function overlap between two $0^+$ states dominates the $E0$ transition. The competition between spherical and deformed configurations results in a small $E0$ transition strength, suggesting shape coexistence in $^{70}\mathrm{Ca}$ [Fig.~\ref{fig6}(a)], $^{78}\mathrm{Ni}$ [Fig.~\ref{fig6}(e)], and $^{80}\mathrm{Zn}$ [Fig.~\ref{fig6}(f)]. By contrast, Figs.~\ref{fig6}(b)--(d) show similar wave-function distributions for the $0^+_1$ and $0^+_2$ states, leading to enhanced $E0$ transitions and indicating considerable shape mixing in $^{72}\mathrm{Ti}$, $^{74}\mathrm{Cr}$, and $^{76}\mathrm{Fe}$.

%
%
%
\section{CONCLUSIONS}
 \label{conclusion}
We apply MR-CDFT with the relativistic EDF PC-PK1 to study low-lying states associated with the structural evolution of neutron-rich $N=50$ isotones from $^{80}\mathrm{Zn}$ to $^{70}\mathrm{Ca}$, and compare the results with those from valence-space and full single-particle-space methods, including shell models, \textit{ab initio} calculations, and EDF-based BMF approaches. Our calculations reproduce the low-lying spectrum of $^{78}\mathrm{Ni}$, yielding $E(2^+_1)=2.64$~MeV and $B(E2;2^+_1\to0^+_1)=106~e^2\text{fm}^4$, consistent with the doubly magic nature of this nucleus. We identify a collective rotational band built on the prolate $0^+_2$ state, with an enhanced intraband $B(E2;2^+_2\to0^+_2)=451~e^{2}\mathrm{fm}^{4}$ and a weak $E0$ transition, $\rho^2(E0;0^+_2\to0^+_1)=2\times10^{-3}$, between the spherical $0^+_1$ and deformed $0^+_2$ states. This suggests weak shape mixing and establishes shape coexistence in $^{78}\mathrm{Ni}$. In contrast, deformation and collectivity develop in $^{76}\mathrm{Fe}$ and $^{74}\mathrm{Cr}$, where the large $E0$ strengths indicate strong shape mixing. These findings provide a unified picture of the structural evolution along the $N=50$ isotonic chain, from the spherical doubly magic nucleus $^{78}\mathrm{Ni}$ toward the more collective $^{76}\mathrm{Fe}$ and $^{74}\mathrm{Cr}$, and motivate future measurements of transition strengths and lifetimes in this key region. Nevertheless, a quantitatively accurate description of extremely neutron-rich nuclei remains challenging because of strong many-body correlations and continuum coupling. The effects of additional deformation degrees of freedom, such as triaxial and octupole deformations, and of continuum coupling on low-lying states are beyond the scope of this work.

%
\section*{ACKNOWLEDGMENTS}
 We thank Jiangming Yao for fruitful discussion. This work was partially supported by the National Natural Science Foundation of China (Nos. 12465020, 12005082), the PhD Foundation of Chongqing Normal University (No.23XLB010), the Science and Technology Research Program of Chongqing Municipal Education Commission (No. KJQN202300509).
%
\providecommand{\noopsort}[1]{}\providecommand{\singleletter}[1]{#1}%

\begin{thebibliography}{68}%
\makeatletter
\providecommand \@ifxundefined [1]{%
 \@ifx{#1\undefined}
}%
\providecommand \@ifnum [1]{%
 \ifnum #1\expandafter \@firstoftwo
 \else \expandafter \@secondoftwo
 \fi
}%
\providecommand \@ifx [1]{%
 \ifx #1\expandafter \@firstoftwo
 \else \expandafter \@secondoftwo
 \fi
}%
\providecommand \natexlab [1]{#1}%
\providecommand \enquote  [1]{``#1''}%
\providecommand \bibnamefont  [1]{#1}%
\providecommand \bibfnamefont [1]{#1}%
\providecommand \citenamefont [1]{#1}%
\providecommand \href@noop [0]{\@secondoftwo}%
\providecommand \href [0]{\begingroup \@sanitize@url \@href}%
\providecommand \@href[1]{\@@startlink{#1}\@@href}%
\providecommand \@@href[1]{\endgroup#1\@@endlink}%
\providecommand \@sanitize@url [0]{\catcode `\\12\catcode `\$12\catcode
  `\&12\catcode `\#12\catcode `\^12\catcode `\_12\catcode `\%12\relax}%
\providecommand \@@startlink[1]{}%
\providecommand \@@endlink[0]{}%
\providecommand \url  [0]{\begingroup\@sanitize@url \@url }%
\providecommand \@url [1]{\endgroup\@href {#1}{\urlprefix }}%
\providecommand \urlprefix  [0]{URL }%
\providecommand \Eprint [0]{\href }%
\providecommand \doibase [0]{https://doi.org/}%
\providecommand \selectlanguage [0]{\@gobble}%
\providecommand \bibinfo  [0]{\@secondoftwo}%
\providecommand \bibfield  [0]{\@secondoftwo}%
\providecommand \translation [1]{[#1]}%
\providecommand \BibitemOpen [0]{}%
\providecommand \bibitemStop [0]{}%
\providecommand \bibitemNoStop [0]{.\EOS\space}%
\providecommand \EOS [0]{\spacefactor3000\relax}%
\providecommand \BibitemShut  [1]{\csname bibitem#1\endcsname}%
\let\auto@bib@innerbib\@empty
\bibitem [{\citenamefont {Erler}\ \emph {et~al.}(2012)\citenamefont {Erler},
  \citenamefont {Birge}, \citenamefont {Kortelainen} \emph {et~al.}}]{Erler12}%
  \BibitemOpen
  \bibfield  {author} {\bibinfo {author} {\bibfnamefont {J.}~\bibnamefont
  {Erler}}, \bibinfo {author} {\bibfnamefont {N.}~\bibnamefont {Birge}},
  \bibinfo {author} {\bibfnamefont {M.}~\bibnamefont {Kortelainen}}, \emph
  {et~al.},\ }\bibfield  {title} {\bibinfo {title} {{The limits of the nuclear
  landscape}},\ }\href {https://doi.org/10.1038/nature11188} {\bibfield
  {journal} {\bibinfo  {journal} {Nature}\ }\textbf {\bibinfo {volume} {486}},\
  \bibinfo {pages} {509} (\bibinfo {year} {2012})}\BibitemShut {NoStop}%
\bibitem [{\citenamefont {Wienholtz}\ \emph {et~al.}(2013)\citenamefont
  {Wienholtz}, \citenamefont {Beck}, \citenamefont {Blaum} \emph
  {et~al.}}]{Wienholtz13}%
  \BibitemOpen
  \bibfield  {author} {\bibinfo {author} {\bibfnamefont {F.}~\bibnamefont
  {Wienholtz}}, \bibinfo {author} {\bibfnamefont {D.}~\bibnamefont {Beck}},
  \bibinfo {author} {\bibfnamefont {K.}~\bibnamefont {Blaum}}, \emph {et~al.},\
  }\bibfield  {title} {\bibinfo {title} {{Masses of exotic calcium isotopes pin
  down nuclear forces}},\ }\href {https://doi.org/10.1038/nature12226}
  {\bibfield  {journal} {\bibinfo  {journal} {Nature}\ }\textbf {\bibinfo
  {volume} {498}},\ \bibinfo {pages} {346} (\bibinfo {year}
  {2013})}\BibitemShut {NoStop}%
\bibitem [{\citenamefont {Ye}\ \emph {et~al.}(2025)\citenamefont {Ye},
  \citenamefont {Yang}, \citenamefont {Sakurai},\ and\ \citenamefont
  {Hu}}]{Ye25}%
  \BibitemOpen
  \bibfield  {author} {\bibinfo {author} {\bibfnamefont {Y.}~\bibnamefont
  {Ye}}, \bibinfo {author} {\bibfnamefont {X.}~\bibnamefont {Yang}}, \bibinfo
  {author} {\bibfnamefont {H.}~\bibnamefont {Sakurai}},\ and\ \bibinfo {author}
  {\bibfnamefont {B.}~\bibnamefont {Hu}},\ }\bibfield  {title} {\bibinfo
  {title} {Physics of exotic nuclei},\ }\href
  {https://doi.org/10.1038/s42254-024-00782-5} {\bibfield  {journal} {\bibinfo
  {journal} {Nat. Rev. Phys.}\ }\textbf {\bibinfo {volume} {7}},\ \bibinfo
  {pages} {21} (\bibinfo {year} {2025})}\BibitemShut {NoStop}%
\bibitem [{\citenamefont {Bender}\ \emph {et~al.}(2003)\citenamefont {Bender},
  \citenamefont {Heenen},\ and\ \citenamefont {Reinhard}}]{Bender03}%
  \BibitemOpen
  \bibfield  {author} {\bibinfo {author} {\bibfnamefont {M.}~\bibnamefont
  {Bender}}, \bibinfo {author} {\bibfnamefont {P.-H.}\ \bibnamefont {Heenen}},\
  and\ \bibinfo {author} {\bibfnamefont {P.-G.}\ \bibnamefont {Reinhard}},\
  }\bibfield  {title} {\bibinfo {title} {Self-consistent mean-field models for
  nuclear structure},\ }\href {https://doi.org/10.1103/RevModPhys.75.121}
  {\bibfield  {journal} {\bibinfo  {journal} {Rev. Mod. Phys.}\ }\textbf
  {\bibinfo {volume} {75}},\ \bibinfo {pages} {121} (\bibinfo {year}
  {2003})}\BibitemShut {NoStop}%
\bibitem [{\citenamefont {Vretenar}\ \emph {et~al.}(2005)\citenamefont
  {Vretenar}, \citenamefont {Afanasjev}, \citenamefont {Lalazissis},\ and\
  \citenamefont {Ring}}]{Vretenar05}%
  \BibitemOpen
  \bibfield  {author} {\bibinfo {author} {\bibfnamefont {D.}~\bibnamefont
  {Vretenar}}, \bibinfo {author} {\bibfnamefont {A.~V.}\ \bibnamefont
  {Afanasjev}}, \bibinfo {author} {\bibfnamefont {G.~A.}\ \bibnamefont
  {Lalazissis}},\ and\ \bibinfo {author} {\bibfnamefont {P.}~\bibnamefont
  {Ring}},\ }\bibfield  {title} {\bibinfo {title} {Relativistic
  {Hartree}–{Bogoliubov} theory: static and dynamic aspects of exotic nuclear
  structure},\ }\href {https://doi.org/10.1016/j.physrep.2004.10.001}
  {\bibfield  {journal} {\bibinfo  {journal} {Phys. Rep.}\ }\textbf {\bibinfo
  {volume} {409}},\ \bibinfo {pages} {101} (\bibinfo {year}
  {2005})}\BibitemShut {NoStop}%
\bibitem [{\citenamefont {Otsuka}\ \emph {et~al.}(2020)\citenamefont {Otsuka},
  \citenamefont {Gade}, \citenamefont {Sorlin}, \citenamefont {Suzuki},\ and\
  \citenamefont {Utsuno}}]{Otsuka20}%
  \BibitemOpen
  \bibfield  {author} {\bibinfo {author} {\bibfnamefont {T.}~\bibnamefont
  {Otsuka}}, \bibinfo {author} {\bibfnamefont {A.}~\bibnamefont {Gade}},
  \bibinfo {author} {\bibfnamefont {O.}~\bibnamefont {Sorlin}}, \bibinfo
  {author} {\bibfnamefont {T.}~\bibnamefont {Suzuki}},\ and\ \bibinfo {author}
  {\bibfnamefont {Y.}~\bibnamefont {Utsuno}},\ }\bibfield  {title} {\bibinfo
  {title} {Evolution of shell structure in exotic nuclei},\ }\href
  {https://doi.org/10.1103/RevModPhys.92.015002} {\bibfield  {journal}
  {\bibinfo  {journal} {Rev. Mod. Phys.}\ }\textbf {\bibinfo {volume} {92}},\
  \bibinfo {pages} {015002} (\bibinfo {year} {2020})}\BibitemShut {NoStop}%
\bibitem [{\citenamefont {Pfeiffer}\ \emph {et~al.}(2001)\citenamefont
  {Pfeiffer}, \citenamefont {Kratz}, \citenamefont {Thielemann},\ and\
  \citenamefont {Walters}}]{Pfeiffer01}%
  \BibitemOpen
  \bibfield  {author} {\bibinfo {author} {\bibfnamefont {B.}~\bibnamefont
  {Pfeiffer}}, \bibinfo {author} {\bibfnamefont {K.-L.}\ \bibnamefont {Kratz}},
  \bibinfo {author} {\bibfnamefont {F.-K.}\ \bibnamefont {Thielemann}},\ and\
  \bibinfo {author} {\bibfnamefont {W.~B.}\ \bibnamefont {Walters}},\
  }\bibfield  {title} {\bibinfo {title} {Nuclear structure studies for the
  astrophysical $r$-process},\ }\href
  {https://doi.org/10.1016/S0375-9474(01)01141-1} {\bibfield  {journal}
  {\bibinfo  {journal} {Nucl. Phys. A}\ }\textbf {\bibinfo {volume} {693}},\
  \bibinfo {pages} {282} (\bibinfo {year} {2001})}\BibitemShut {NoStop}%
\bibitem [{\citenamefont {Arnould}\ \emph {et~al.}(2007)\citenamefont
  {Arnould}, \citenamefont {Goriely},\ and\ \citenamefont
  {Takahashi}}]{Arnould07}%
  \BibitemOpen
  \bibfield  {author} {\bibinfo {author} {\bibfnamefont {M.}~\bibnamefont
  {Arnould}}, \bibinfo {author} {\bibfnamefont {S.}~\bibnamefont {Goriely}},\
  and\ \bibinfo {author} {\bibfnamefont {K.}~\bibnamefont {Takahashi}},\
  }\bibfield  {title} {\bibinfo {title} {The r-process of stellar
  nucleosynthesis: Astrophysics and nuclear physics achievements and
  mysteries},\ }\href {https://doi.org/10.1016/j.physrep.2007.06.002}
  {\bibfield  {journal} {\bibinfo  {journal} {Phys. Rep.}\ }\textbf {\bibinfo
  {volume} {450}},\ \bibinfo {pages} {97} (\bibinfo {year} {2007})}\BibitemShut
  {NoStop}%
\bibitem [{\citenamefont {Baruah}\ \emph {et~al.}(2008)\citenamefont {Baruah},
  \citenamefont {Audi}, \citenamefont {Blaum} \emph {et~al.}}]{Baruah2008}%
  \BibitemOpen
  \bibfield  {author} {\bibinfo {author} {\bibfnamefont {S.}~\bibnamefont
  {Baruah}}, \bibinfo {author} {\bibfnamefont {G.}~\bibnamefont {Audi}},
  \bibinfo {author} {\bibfnamefont {K.}~\bibnamefont {Blaum}}, \emph {et~al.},\
  }\bibfield  {title} {\bibinfo {title} {Mass measurements beyond the major
  $r$-process waiting point $^{80}${Zn}},\ }\href
  {https://doi.org/10.1103/PhysRevLett.101.262501} {\bibfield  {journal}
  {\bibinfo  {journal} {Phys. Rev. Lett.}\ }\textbf {\bibinfo {volume} {101}},\
  \bibinfo {pages} {262501} (\bibinfo {year} {2008})}\BibitemShut {NoStop}%
\bibitem [{\citenamefont {Cowan}\ \emph {et~al.}(2021)\citenamefont {Cowan},
  \citenamefont {Sneden}, \citenamefont {Lawler} \emph {et~al.}}]{Cowan21}%
  \BibitemOpen
  \bibfield  {author} {\bibinfo {author} {\bibfnamefont {J.~J.}\ \bibnamefont
  {Cowan}}, \bibinfo {author} {\bibfnamefont {C.}~\bibnamefont {Sneden}},
  \bibinfo {author} {\bibfnamefont {J.~E.}\ \bibnamefont {Lawler}}, \emph
  {et~al.},\ }\bibfield  {title} {\bibinfo {title} {Origin of the heaviest
  elements: The rapid neutron-capture process},\ }\href
  {https://doi.org/10.1103/RevModPhys.93.015002} {\bibfield  {journal}
  {\bibinfo  {journal} {Rev. Mod. Phys.}\ }\textbf {\bibinfo {volume} {93}},\
  \bibinfo {pages} {015002} (\bibinfo {year} {2021})}\BibitemShut {NoStop}%
\bibitem [{\citenamefont {Bastin}\ \emph {et~al.}(2007)\citenamefont {Bastin},
  \citenamefont {Gr{\'e}vy}, \citenamefont {Sohler} \emph {et~al.}}]{Bastin07}%
  \BibitemOpen
  \bibfield  {author} {\bibinfo {author} {\bibfnamefont {B.}~\bibnamefont
  {Bastin}}, \bibinfo {author} {\bibfnamefont {S.}~\bibnamefont {Gr{\'e}vy}},
  \bibinfo {author} {\bibfnamefont {D.}~\bibnamefont {Sohler}}, \emph
  {et~al.},\ }\bibfield  {title} {\bibinfo {title} {Collapse of the {$N=28$}
  shell closure in {$^{42}\mathrm{Si}$}},\ }\href
  {https://doi.org/10.1103/PhysRevLett.99.022503} {\bibfield  {journal}
  {\bibinfo  {journal} {Phys. Rev. Lett.}\ }\textbf {\bibinfo {volume} {99}},\
  \bibinfo {pages} {022503} (\bibinfo {year} {2007})}\BibitemShut {NoStop}%
\bibitem [{\citenamefont {Ding}\ \emph {et~al.}(2026)\citenamefont {Ding},
  \citenamefont {Wang}, \citenamefont {Yao} \emph {et~al.}}]{Ding26}%
  \BibitemOpen
  \bibfield  {author} {\bibinfo {author} {\bibfnamefont {C.~R.}\ \bibnamefont
  {Ding}}, \bibinfo {author} {\bibfnamefont {C.~C.}\ \bibnamefont {Wang}},
  \bibinfo {author} {\bibfnamefont {J.~M.}\ \bibnamefont {Yao}}, \emph
  {et~al.},\ }\bibfield  {title} {\bibinfo {title} {From spin to pseudospin
  symmetry: The origin of magic numbers in nuclear structure},\ }\href
  {https://doi.org/10.1103/8lzc-j1lx} {\bibfield  {journal} {\bibinfo
  {journal} {Phys. Rev. Lett.}\ }\textbf {\bibinfo {volume} {136}},\ \bibinfo
  {pages} {052501} (\bibinfo {year} {2026})}\BibitemShut {NoStop}%
\bibitem [{\citenamefont {Gray}\ \emph {et~al.}(2023)\citenamefont {Gray},
  \citenamefont {Allmond}, \citenamefont {Xu} \emph {et~al.}}]{Gray23}%
  \BibitemOpen
  \bibfield  {author} {\bibinfo {author} {\bibfnamefont {T.~J.}\ \bibnamefont
  {Gray}}, \bibinfo {author} {\bibfnamefont {J.~M.}\ \bibnamefont {Allmond}},
  \bibinfo {author} {\bibfnamefont {Z.}~\bibnamefont {Xu}}, \emph {et~al.},\
  }\bibfield  {title} {\bibinfo {title} {Microsecond isomer at the {$N=20$}
  island of shape inversion observed at frib},\ }\href
  {https://doi.org/10.1103/PhysRevLett.130.242501} {\bibfield  {journal}
  {\bibinfo  {journal} {Phys. Rev. Lett.}\ }\textbf {\bibinfo {volume} {130}},\
  \bibinfo {pages} {242501} (\bibinfo {year} {2023})}\BibitemShut {NoStop}%
\bibitem [{\citenamefont {Rocchini}\ \emph {et~al.}(2023)\citenamefont
  {Rocchini}, \citenamefont {Garrett}, \citenamefont
  {Zieli\ifmmode~\acute{n}\else \'{n}\fi{}ska} \emph {et~al.}}]{Rocchini23}%
  \BibitemOpen
  \bibfield  {author} {\bibinfo {author} {\bibfnamefont {M.}~\bibnamefont
  {Rocchini}}, \bibinfo {author} {\bibfnamefont {P.~E.}\ \bibnamefont
  {Garrett}}, \bibinfo {author} {\bibfnamefont {M.}~\bibnamefont
  {Zieli\ifmmode~\acute{n}\else \'{n}\fi{}ska}}, \emph {et~al.},\ }\bibfield
  {title} {\bibinfo {title} {First evidence of axial shape asymmetry and
  configuration coexistence in $^{74}\mathrm{Zn}$: Suggestion for a northern
  extension of the {$N=40$} island of inversion},\ }\href
  {https://doi.org/10.1103/PhysRevLett.130.122502} {\bibfield  {journal}
  {\bibinfo  {journal} {Phys. Rev. Lett.}\ }\textbf {\bibinfo {volume} {130}},\
  \bibinfo {pages} {122502} (\bibinfo {year} {2023})}\BibitemShut {NoStop}%
\bibitem [{\citenamefont {Zhou}\ \emph {et~al.}(2025)\citenamefont {Zhou},
  \citenamefont {Ding}, \citenamefont {Yao} \emph {et~al.}}]{Zhou25}%
  \BibitemOpen
  \bibfield  {author} {\bibinfo {author} {\bibfnamefont {E.~F.}\ \bibnamefont
  {Zhou}}, \bibinfo {author} {\bibfnamefont {C.~R.}\ \bibnamefont {Ding}},
  \bibinfo {author} {\bibfnamefont {J.~M.}\ \bibnamefont {Yao}}, \emph
  {et~al.},\ }\bibfield  {title} {\bibinfo {title} {Ab initio nuclear shape
  coexistence and emergence of island of inversion around {$N=20$}},\ }\href
  {https://doi.org/10.1016/j.physletb.2025.139464} {\bibfield  {journal}
  {\bibinfo  {journal} {Phys. Lett. B}\ }\textbf {\bibinfo {volume} {865}},\
  \bibinfo {pages} {139464} (\bibinfo {year} {2025})}\BibitemShut {NoStop}%
\bibitem [{\citenamefont {Wood}\ \emph {et~al.}(1992)\citenamefont {Wood},
  \citenamefont {Heyde}, \citenamefont {Nazarewicz}, \citenamefont {Huyse},\
  and\ \citenamefont {van Duppen}}]{Wood92}%
  \BibitemOpen
  \bibfield  {author} {\bibinfo {author} {\bibfnamefont {J.~L.}\ \bibnamefont
  {Wood}}, \bibinfo {author} {\bibfnamefont {K.}~\bibnamefont {Heyde}},
  \bibinfo {author} {\bibfnamefont {W.}~\bibnamefont {Nazarewicz}}, \bibinfo
  {author} {\bibfnamefont {M.}~\bibnamefont {Huyse}},\ and\ \bibinfo {author}
  {\bibfnamefont {P.}~\bibnamefont {van Duppen}},\ }\bibfield  {title}
  {\bibinfo {title} {Coexistence in even-mass nuclei},\ }\href
  {https://doi.org/doi.org/10.1016/0370-1573(92)90095-H} {\bibfield  {journal}
  {\bibinfo  {journal} {Phys. Rep.}\ }\textbf {\bibinfo {volume} {215}},\
  \bibinfo {pages} {101} (\bibinfo {year} {1992})}\BibitemShut {NoStop}%
\bibitem [{\citenamefont {Heyde}\ and\ \citenamefont {Wood}(2011)}]{Heyde11}%
  \BibitemOpen
  \bibfield  {author} {\bibinfo {author} {\bibfnamefont {K.}~\bibnamefont
  {Heyde}}\ and\ \bibinfo {author} {\bibfnamefont {J.~L.}\ \bibnamefont
  {Wood}},\ }\bibfield  {title} {\bibinfo {title} {Shape coexistence in atomic
  nuclei},\ }\href {https://doi.org/10.1103/RevModPhys.83.1467} {\bibfield
  {journal} {\bibinfo  {journal} {Rev. Mod. Phys.}\ }\textbf {\bibinfo {volume}
  {83}},\ \bibinfo {pages} {1467} (\bibinfo {year} {2011})}\BibitemShut
  {NoStop}%
\bibitem [{\citenamefont {Orlandi}\ \emph {et~al.}(2015)\citenamefont
  {Orlandi}, \citenamefont {Mücher}, \citenamefont {Raabe} \emph
  {et~al.}}]{Orlandi15}%
  \BibitemOpen
  \bibfield  {author} {\bibinfo {author} {\bibfnamefont {R.}~\bibnamefont
  {Orlandi}}, \bibinfo {author} {\bibfnamefont {D.}~\bibnamefont {Mücher}},
  \bibinfo {author} {\bibfnamefont {R.}~\bibnamefont {Raabe}}, \emph {et~al.},\
  }\bibfield  {title} {\bibinfo {title} {Single-neutron orbits near
  $^{78}${Ni}: Spectroscopy of the {$N=49$} isotope $^{79}${Zn}},\ }\href
  {https://doi.org/10.1016/j.physletb.2014.12.006} {\bibfield  {journal}
  {\bibinfo  {journal} {Phys. Lett. B}\ }\textbf {\bibinfo {volume} {740}},\
  \bibinfo {pages} {298} (\bibinfo {year} {2015})}\BibitemShut {NoStop}%
\bibitem [{\citenamefont {Welker}\ \emph {et~al.}(2017)\citenamefont {Welker},
  \citenamefont {Althubiti}, \citenamefont {Atanasov} \emph
  {et~al.}}]{Welker17}%
  \BibitemOpen
  \bibfield  {author} {\bibinfo {author} {\bibfnamefont {A.}~\bibnamefont
  {Welker}}, \bibinfo {author} {\bibfnamefont {N.~A.~S.}\ \bibnamefont
  {Althubiti}}, \bibinfo {author} {\bibfnamefont {D.}~\bibnamefont {Atanasov}},
  \emph {et~al.},\ }\bibfield  {title} {\bibinfo {title} {Binding energy of
  $^{79}${Cu}: Probing the structure of the doubly magic $^{78}${Ni} from only
  one proton away},\ }\href {https://doi.org/10.1103/PhysRevLett.119.192502}
  {\bibfield  {journal} {\bibinfo  {journal} {Phys. Rev. Lett.}\ }\textbf
  {\bibinfo {volume} {119}},\ \bibinfo {pages} {192502} (\bibinfo {year}
  {2017})}\BibitemShut {NoStop}%
\bibitem [{\citenamefont {Vajta}\ \emph {et~al.}(2018)\citenamefont {Vajta},
  \citenamefont {Sohler}, \citenamefont {Shiga} \emph {et~al.}}]{Vajta18}%
  \BibitemOpen
  \bibfield  {author} {\bibinfo {author} {\bibfnamefont {Z.}~\bibnamefont
  {Vajta}}, \bibinfo {author} {\bibfnamefont {D.}~\bibnamefont {Sohler}},
  \bibinfo {author} {\bibfnamefont {Y.}~\bibnamefont {Shiga}}, \emph {et~al.},\
  }\bibfield  {title} {\bibinfo {title} {Proton single particle energies next
  to $^{78}${Ni}: Spectroscopy of $^{77}${Cu} via single proton knock-out
  reaction},\ }\href {https://doi.org/10.1016/j.physletb.2018.05.023}
  {\bibfield  {journal} {\bibinfo  {journal} {Phys. Lett. B}\ }\textbf
  {\bibinfo {volume} {782}},\ \bibinfo {pages} {99} (\bibinfo {year}
  {2018})}\BibitemShut {NoStop}%
\bibitem [{\citenamefont {Olivier}\ \emph {et~al.}(2017)\citenamefont
  {Olivier}, \citenamefont {Franchoo}, \citenamefont {Niikura} \emph
  {et~al.}}]{Olivier17}%
  \BibitemOpen
  \bibfield  {author} {\bibinfo {author} {\bibfnamefont {L.}~\bibnamefont
  {Olivier}}, \bibinfo {author} {\bibfnamefont {S.}~\bibnamefont {Franchoo}},
  \bibinfo {author} {\bibfnamefont {M.}~\bibnamefont {Niikura}}, \emph
  {et~al.},\ }\bibfield  {title} {\bibinfo {title} {Persistence of the {$Z=28$}
  shell gap around $^{78}${Ni}: first spectroscopy of $^{79}${Cu}},\ }\href
  {https://doi.org/10.1103/PhysRevLett.119.192501} {\bibfield  {journal}
  {\bibinfo  {journal} {Phys. Rev. Lett.}\ }\textbf {\bibinfo {volume} {119}},\
  \bibinfo {pages} {192501} (\bibinfo {year} {2017})}\BibitemShut {NoStop}%
\bibitem [{\citenamefont {Nies}\ \emph {et~al.}(2023)\citenamefont {Nies},
  \citenamefont {Canete}, \citenamefont {Dao} \emph {et~al.}}]{Nies23}%
  \BibitemOpen
  \bibfield  {author} {\bibinfo {author} {\bibfnamefont {L.}~\bibnamefont
  {Nies}}, \bibinfo {author} {\bibfnamefont {L.}~\bibnamefont {Canete}},
  \bibinfo {author} {\bibfnamefont {D.~D.}\ \bibnamefont {Dao}}, \emph
  {et~al.},\ }\bibfield  {title} {\bibinfo {title} {Further evidence for shape
  coexistence in $^{79}${Zn} near doubly magic $^{78}${Ni}},\ }\href
  {https://doi.org/10.1103/PhysRevLett.131.222503} {\bibfield  {journal}
  {\bibinfo  {journal} {Phys. Rev. Lett.}\ }\textbf {\bibinfo {volume} {131}},\
  \bibinfo {pages} {222503} (\bibinfo {year} {2023})}\BibitemShut {NoStop}%
\bibitem [{\citenamefont {Taniuchi}\ \emph {et~al.}(2019)\citenamefont
  {Taniuchi}, \citenamefont {Santamaria}, \citenamefont {Doornenbal} \emph
  {et~al.}}]{Taniuchi2019}%
  \BibitemOpen
  \bibfield  {author} {\bibinfo {author} {\bibfnamefont {R.}~\bibnamefont
  {Taniuchi}}, \bibinfo {author} {\bibfnamefont {C.}~\bibnamefont
  {Santamaria}}, \bibinfo {author} {\bibfnamefont {P.}~\bibnamefont
  {Doornenbal}}, \emph {et~al.},\ }\bibfield  {title} {\bibinfo {title}
  {$^{78}$$\mathrm{Ni}$ revealed as a doubly magic stronghold against nuclear
  deformation},\ }\href {https://doi.org/10.1038/s41586-019-1155-x} {\bibfield
  {journal} {\bibinfo  {journal} {Nature}\ }\textbf {\bibinfo {volume} {569}},\
  \bibinfo {pages} {53} (\bibinfo {year} {2019})}\BibitemShut {NoStop}%
\bibitem [{\citenamefont {Gottardo}\ \emph {et~al.}(2016)\citenamefont
  {Gottardo}, \citenamefont {Verney}, \citenamefont {Delafosse} \emph
  {et~al.}}]{Gottardo16}%
  \BibitemOpen
  \bibfield  {author} {\bibinfo {author} {\bibfnamefont {A.}~\bibnamefont
  {Gottardo}}, \bibinfo {author} {\bibfnamefont {D.}~\bibnamefont {Verney}},
  \bibinfo {author} {\bibfnamefont {C.}~\bibnamefont {Delafosse}}, \emph
  {et~al.},\ }\bibfield  {title} {\bibinfo {title} {First evidence of shape
  coexistence in the $^{78}\mathrm{Ni}$ region: Intruder ${0}_{2}^{+}$ state in
  $^{80}\mathrm{Ge}$},\ }\href {https://doi.org/10.1103/PhysRevLett.116.182501}
  {\bibfield  {journal} {\bibinfo  {journal} {Phys. Rev. Lett.}\ }\textbf
  {\bibinfo {volume} {116}},\ \bibinfo {pages} {182501} (\bibinfo {year}
  {2016})}\BibitemShut {NoStop}%
\bibitem [{\citenamefont {Yang}\ \emph {et~al.}(2016)\citenamefont {Yang},
  \citenamefont {Wraith}, \citenamefont {Xie} \emph {et~al.}}]{Yang16}%
  \BibitemOpen
  \bibfield  {author} {\bibinfo {author} {\bibfnamefont {X.~F.}\ \bibnamefont
  {Yang}}, \bibinfo {author} {\bibfnamefont {C.}~\bibnamefont {Wraith}},
  \bibinfo {author} {\bibfnamefont {L.}~\bibnamefont {Xie}}, \emph {et~al.},\
  }\bibfield  {title} {\bibinfo {title} {Isomer shift and magnetic moment of
  the long-lived 1/2$^{+}$ isomer in $^{79}_{30}${Zn}$_{49}$: Signature of
  shape coexistence near $^{78}${Ni}},\ }\href
  {https://doi.org/10.1103/PhysRevLett.116.182502} {\bibfield  {journal}
  {\bibinfo  {journal} {Phys. Rev. Lett.}\ }\textbf {\bibinfo {volume} {116}},\
  \bibinfo {pages} {182502} (\bibinfo {year} {2016})}\BibitemShut {NoStop}%
\bibitem [{\citenamefont {Nowacki}\ \emph {et~al.}(2021)\citenamefont
  {Nowacki}, \citenamefont {Obertelli},\ and\ \citenamefont
  {Poves}}]{Nowacki21}%
  \BibitemOpen
  \bibfield  {author} {\bibinfo {author} {\bibfnamefont {F.}~\bibnamefont
  {Nowacki}}, \bibinfo {author} {\bibfnamefont {A.}~\bibnamefont {Obertelli}},\
  and\ \bibinfo {author} {\bibfnamefont {A.}~\bibnamefont {Poves}},\ }\bibfield
   {title} {\bibinfo {title} {The neutron-rich edge of the nuclear landscape:
  Experiment and theory},\ }\href {https://doi.org/10.1016/j.ppnp.2021.103866}
  {\bibfield  {journal} {\bibinfo  {journal} {Prog. Part. Nucl. Phys.}\
  }\textbf {\bibinfo {volume} {120}},\ \bibinfo {pages} {103866} (\bibinfo
  {year} {2021})}\BibitemShut {NoStop}%
\bibitem [{\citenamefont {Taniuchi}(2026)}]{Taniuchi26}%
  \BibitemOpen
  \bibfield  {author} {\bibinfo {author} {\bibfnamefont {R.}~\bibnamefont
  {Taniuchi}},\ }\bibfield  {title} {\bibinfo {title} {Competition of the
  {Shell} {Closure} and {Deformations} around the {Doubly} {Magic}
  $^{78}\mathrm{Ni}$},\ }\href {https://doi.org/10.1093/ptep/ptaf084}
  {\bibfield  {journal} {\bibinfo  {journal} {Prog. Theor. Exp. Phys.}\
  }\textbf {\bibinfo {volume} {2026}},\ \bibinfo {pages} {04A103} (\bibinfo
  {year} {2026})}\BibitemShut {NoStop}%
\bibitem [{\citenamefont {Nowacki}\ \emph {et~al.}(2016)\citenamefont
  {Nowacki}, \citenamefont {Poves}, \citenamefont {Caurier},\ and\
  \citenamefont {Bounthong}}]{Nowacki2016}%
  \BibitemOpen
  \bibfield  {author} {\bibinfo {author} {\bibfnamefont {F.}~\bibnamefont
  {Nowacki}}, \bibinfo {author} {\bibfnamefont {A.}~\bibnamefont {Poves}},
  \bibinfo {author} {\bibfnamefont {E.}~\bibnamefont {Caurier}},\ and\ \bibinfo
  {author} {\bibfnamefont {B.}~\bibnamefont {Bounthong}},\ }\bibfield  {title}
  {\bibinfo {title} {Shape coexistence in$^{78}$$\mathrm{Ni}$ as the portal to
  the fifth island of inversion},\ }\href
  {https://doi.org/10.1103/PhysRevLett.117.272501} {\bibfield  {journal}
  {\bibinfo  {journal} {Phys. Rev. Lett.}\ }\textbf {\bibinfo {volume} {117}},\
  \bibinfo {pages} {272501} (\bibinfo {year} {2016})}\BibitemShut {NoStop}%
\bibitem [{\citenamefont {Tsunoda}\ \emph {et~al.}(2014)\citenamefont
  {Tsunoda}, \citenamefont {Otsuka}, \citenamefont {Shimizu}, \citenamefont
  {Honma},\ and\ \citenamefont {Utsuno}}]{Tsunoda2014}%
  \BibitemOpen
  \bibfield  {author} {\bibinfo {author} {\bibfnamefont {Y.}~\bibnamefont
  {Tsunoda}}, \bibinfo {author} {\bibfnamefont {T.}~\bibnamefont {Otsuka}},
  \bibinfo {author} {\bibfnamefont {N.}~\bibnamefont {Shimizu}}, \bibinfo
  {author} {\bibfnamefont {M.}~\bibnamefont {Honma}},\ and\ \bibinfo {author}
  {\bibfnamefont {Y.}~\bibnamefont {Utsuno}},\ }\bibfield  {title} {\bibinfo
  {title} {{Novel shape evolution in exotic Ni isotopes and
  configuration-dependent shell structure}},\ }\href
  {https://doi.org/10.1103/PhysRevC.89.031301} {\bibfield  {journal} {\bibinfo
  {journal} {Phys. Rev. C}\ }\textbf {\bibinfo {volume} {89}},\ \bibinfo
  {pages} {031301} (\bibinfo {year} {2014})}\BibitemShut {NoStop}%
\bibitem [{\citenamefont {Kumar}\ \emph {et~al.}(2026)\citenamefont {Kumar},
  \citenamefont {Shimizu}, \citenamefont {Miyagi}, \citenamefont {Tsunoda},\
  and\ \citenamefont {Utsuno}}]{Kumar26}%
  \BibitemOpen
  \bibfield  {author} {\bibinfo {author} {\bibfnamefont {A.}~\bibnamefont
  {Kumar}}, \bibinfo {author} {\bibfnamefont {N.}~\bibnamefont {Shimizu}},
  \bibinfo {author} {\bibfnamefont {T.}~\bibnamefont {Miyagi}}, \bibinfo
  {author} {\bibfnamefont {Y.}~\bibnamefont {Tsunoda}},\ and\ \bibinfo {author}
  {\bibfnamefont {Y.}~\bibnamefont {Utsuno}},\ }\bibfield  {title} {\bibinfo
  {title} {Probing the shape evolution and shell structures in neutron-rich
  {$N=50$} nuclei},\ }\href {https://doi.org/10.1016/j.physletb.2026.140184}
  {\bibfield  {journal} {\bibinfo  {journal} {Phys. Lett. B}\ }\textbf
  {\bibinfo {volume} {874}},\ \bibinfo {pages} {140184} (\bibinfo {year}
  {2026})}\BibitemShut {NoStop}%
\bibitem [{\citenamefont {Li}(2023)}]{Li23}%
  \BibitemOpen
  \bibfield  {author} {\bibinfo {author} {\bibfnamefont {J.~G.}\ \bibnamefont
  {Li}},\ }\bibfield  {title} {\bibinfo {title} {Merging of the island of
  inversion at {$N=40$} and {$N=50$}},\ }\href
  {https://doi.org/10.1016/j.physletb.2023.137893} {\bibfield  {journal}
  {\bibinfo  {journal} {Phys. Lett. B}\ }\textbf {\bibinfo {volume} {840}},\
  \bibinfo {pages} {137893} (\bibinfo {year} {2023})}\BibitemShut {NoStop}%
\bibitem [{\citenamefont {Tichai}\ \emph {et~al.}(2024)\citenamefont {Tichai},
  \citenamefont {Kapás}, \citenamefont {Miyagi} \emph {et~al.}}]{Tichai24}%
  \BibitemOpen
  \bibfield  {author} {\bibinfo {author} {\bibfnamefont {A.}~\bibnamefont
  {Tichai}}, \bibinfo {author} {\bibfnamefont {K.}~\bibnamefont {Kapás}},
  \bibinfo {author} {\bibfnamefont {T.}~\bibnamefont {Miyagi}}, \emph
  {et~al.},\ }\bibfield  {title} {\bibinfo {title} {Spectroscopy of {$N=50$}
  isotones with the valence-space density matrix renormalization group},\
  }\href {https://doi.org/10.1016/j.physletb.2024.138841} {\bibfield  {journal}
  {\bibinfo  {journal} {Phys. Lett. B}\ }\textbf {\bibinfo {volume} {855}},\
  \bibinfo {pages} {138841} (\bibinfo {year} {2024})}\BibitemShut {NoStop}%
\bibitem [{\citenamefont {Hagen}\ \emph {et~al.}(2016)\citenamefont {Hagen},
  \citenamefont {Jansen},\ and\ \citenamefont {Papenbrock}}]{Hagen2016}%
  \BibitemOpen
  \bibfield  {author} {\bibinfo {author} {\bibfnamefont {G.}~\bibnamefont
  {Hagen}}, \bibinfo {author} {\bibfnamefont {G.~R.}\ \bibnamefont {Jansen}},\
  and\ \bibinfo {author} {\bibfnamefont {T.}~\bibnamefont {Papenbrock}},\
  }\bibfield  {title} {\bibinfo {title} {Structure of $^{78}$$\mathrm{Ni}$ from
  first-principles computations},\ }\href
  {https://doi.org/10.1103/PhysRevLett.117.172501} {\bibfield  {journal}
  {\bibinfo  {journal} {Phys. Rev. Lett.}\ }\textbf {\bibinfo {volume} {117}},\
  \bibinfo {pages} {172501} (\bibinfo {year} {2016})}\BibitemShut {NoStop}%
\bibitem [{\citenamefont {Hu}\ \emph {et~al.}(2024)\citenamefont {Hu},
  \citenamefont {Sun}, \citenamefont {Hagen}, \citenamefont {Jansen},\ and\
  \citenamefont {Papenbrock}}]{Hu2024}%
  \BibitemOpen
  \bibfield  {author} {\bibinfo {author} {\bibfnamefont {B.~S.}\ \bibnamefont
  {Hu}}, \bibinfo {author} {\bibfnamefont {Z.~H.}\ \bibnamefont {Sun}},
  \bibinfo {author} {\bibfnamefont {G.}~\bibnamefont {Hagen}}, \bibinfo
  {author} {\bibfnamefont {G.~R.}\ \bibnamefont {Jansen}},\ and\ \bibinfo
  {author} {\bibfnamefont {T.}~\bibnamefont {Papenbrock}},\ }\bibfield  {title}
  {\bibinfo {title} {Ab initio computations from $^{78}$$\mathrm{Ni}$ towards
  $^{70}$$\mathrm{Ca}$ along neutron number {$N=50$}},\ }\href
  {https://doi.org/10.1016/j.physletb.2024.139010} {\bibfield  {journal}
  {\bibinfo  {journal} {Phys. Lett. B}\ }\textbf {\bibinfo {volume} {858}},\
  \bibinfo {pages} {139010} (\bibinfo {year} {2024})}\BibitemShut {NoStop}%
\bibitem [{\citenamefont {Simonis}\ \emph {et~al.}(2017)\citenamefont
  {Simonis}, \citenamefont {Stroberg}, \citenamefont {Hebeler}, \citenamefont
  {Holt},\ and\ \citenamefont {Schwenk}}]{Simonis17}%
  \BibitemOpen
  \bibfield  {author} {\bibinfo {author} {\bibfnamefont {J.}~\bibnamefont
  {Simonis}}, \bibinfo {author} {\bibfnamefont {S.~R.}\ \bibnamefont
  {Stroberg}}, \bibinfo {author} {\bibfnamefont {K.}~\bibnamefont {Hebeler}},
  \bibinfo {author} {\bibfnamefont {J.~D.}\ \bibnamefont {Holt}},\ and\
  \bibinfo {author} {\bibfnamefont {A.}~\bibnamefont {Schwenk}},\ }\bibfield
  {title} {\bibinfo {title} {Saturation with chiral interactions and
  consequences for finite nuclei},\ }\href
  {https://doi.org/10.1103/PhysRevC.96.014303} {\bibfield  {journal} {\bibinfo
  {journal} {Phys. Rev. C}\ }\textbf {\bibinfo {volume} {96}},\ \bibinfo
  {pages} {014303} (\bibinfo {year} {2017})}\BibitemShut {NoStop}%
\bibitem [{\citenamefont {Delaroche}\ \emph {et~al.}(2010)\citenamefont
  {Delaroche}, \citenamefont {Girod}, \citenamefont {Libert} \emph
  {et~al.}}]{Delaroche10}%
  \BibitemOpen
  \bibfield  {author} {\bibinfo {author} {\bibfnamefont {J.~P.}\ \bibnamefont
  {Delaroche}}, \bibinfo {author} {\bibfnamefont {M.}~\bibnamefont {Girod}},
  \bibinfo {author} {\bibfnamefont {J.}~\bibnamefont {Libert}}, \emph
  {et~al.},\ }\bibfield  {title} {\bibinfo {title} {{Structure of even-even
  nuclei using a mapped collective Hamiltonian and the D1S Gogny
  interaction}},\ }\href {https://doi.org/10.1103/PhysRevC.81.014303}
  {\bibfield  {journal} {\bibinfo  {journal} {Phys. Rev. C}\ }\textbf {\bibinfo
  {volume} {81}},\ \bibinfo {pages} {014303} (\bibinfo {year} {2010})},\
  \bibinfo {note} {\url{https://www-phynu.cea.fr/}}\BibitemShut {NoStop}%
\bibitem [{\citenamefont {Péru}\ and\ \citenamefont
  {Goutte}(2008)}]{Peru2008}%
  \BibitemOpen
  \bibfield  {author} {\bibinfo {author} {\bibfnamefont {S.}~\bibnamefont
  {Péru}}\ and\ \bibinfo {author} {\bibfnamefont {H.}~\bibnamefont {Goutte}},\
  }\bibfield  {title} {\bibinfo {title} {Role of deformation on giant
  resonances within the quasiparticle random-phase approximation and the gogny
  force},\ }\href {https://doi.org/10.1103/PhysRevC.77.044313} {\bibfield
  {journal} {\bibinfo  {journal} {Phys. Rev. C}\ }\textbf {\bibinfo {volume}
  {77}},\ \bibinfo {pages} {044313} (\bibinfo {year} {2008})}\BibitemShut
  {NoStop}%
\bibitem [{\citenamefont {Girod}\ \emph {et~al.}(1988)\citenamefont {Girod},
  \citenamefont {{Ph. Dessagne}}, \citenamefont {Bernas} \emph
  {et~al.}}]{Girod88}%
  \BibitemOpen
  \bibfield  {author} {\bibinfo {author} {\bibfnamefont {M.}~\bibnamefont
  {Girod}}, \bibinfo {author} {\bibnamefont {{Ph. Dessagne}}}, \bibinfo
  {author} {\bibfnamefont {M.}~\bibnamefont {Bernas}}, \emph {et~al.},\
  }\bibfield  {title} {\bibinfo {title} {Spectroscopy of neutron-rich nickel
  isotopes: Experimental results and microscopic interpretation},\ }\href
  {https://doi.org/doi.org/10.1103/PhysRevC.37.2600} {\bibfield  {journal}
  {\bibinfo  {journal} {Phys. Rev. C}\ }\textbf {\bibinfo {volume} {37}},\
  \bibinfo {pages} {2600} (\bibinfo {year} {1988})}\BibitemShut {NoStop}%
\bibitem [{\citenamefont {Yao}\ \emph {et~al.}(2009)\citenamefont {Yao},
  \citenamefont {Meng}, \citenamefont {Ring},\ and\ \citenamefont
  {Arteaga}}]{Yao09}%
  \BibitemOpen
  \bibfield  {author} {\bibinfo {author} {\bibfnamefont {J.~M.}\ \bibnamefont
  {Yao}}, \bibinfo {author} {\bibfnamefont {J.}~\bibnamefont {Meng}}, \bibinfo
  {author} {\bibfnamefont {P.}~\bibnamefont {Ring}},\ and\ \bibinfo {author}
  {\bibfnamefont {D.~P.}\ \bibnamefont {Arteaga}},\ }\bibfield  {title}
  {\bibinfo {title} {Three-dimensional angular momentum projection in
  relativistic mean-field theory},\ }\href
  {https://doi.org/10.1103/PhysRevC.79.044312} {\bibfield  {journal} {\bibinfo
  {journal} {Phys. Rev. C}\ }\textbf {\bibinfo {volume} {79}},\ \bibinfo
  {pages} {044312} (\bibinfo {year} {2009})}\BibitemShut {NoStop}%
\bibitem [{\citenamefont {Yao}\ \emph {et~al.}(2010)\citenamefont {Yao},
  \citenamefont {Meng}, \citenamefont {Ring},\ and\ \citenamefont
  {Vretenar}}]{Yao10}%
  \BibitemOpen
  \bibfield  {author} {\bibinfo {author} {\bibfnamefont {J.~M.}\ \bibnamefont
  {Yao}}, \bibinfo {author} {\bibfnamefont {J.}~\bibnamefont {Meng}}, \bibinfo
  {author} {\bibfnamefont {P.}~\bibnamefont {Ring}},\ and\ \bibinfo {author}
  {\bibfnamefont {D.}~\bibnamefont {Vretenar}},\ }\bibfield  {title} {\bibinfo
  {title} {Configuration mixing of angular-momentum-projected triaxial
  relativistic mean-field wave functions},\ }\href
  {https://doi.org/10.1103/PhysRevC.81.044311} {\bibfield  {journal} {\bibinfo
  {journal} {Phys. Rev. C}\ }\textbf {\bibinfo {volume} {81}},\ \bibinfo
  {pages} {044311} (\bibinfo {year} {2010})}\BibitemShut {NoStop}%
\bibitem [{\citenamefont {Wu}\ and\ \citenamefont {Zhou}(2015)}]{Wu15}%
  \BibitemOpen
  \bibfield  {author} {\bibinfo {author} {\bibfnamefont {X.-Y.}\ \bibnamefont
  {Wu}}\ and\ \bibinfo {author} {\bibfnamefont {X.-R.}\ \bibnamefont {Zhou}},\
  }\bibfield  {title} {\bibinfo {title} {Global performance of multireference
  density functional theory for low-lying states in $sd$-shell nuclei},\ }\href
  {https://doi.org/10.1103/PhysRevC.92.054321} {\bibfield  {journal} {\bibinfo
  {journal} {Phys. Rev. C}\ }\textbf {\bibinfo {volume} {92}},\ \bibinfo
  {pages} {054321} (\bibinfo {year} {2015})}\BibitemShut {NoStop}%
\bibitem [{\citenamefont {Yao}(2020)}]{Yao20}%
  \BibitemOpen
  \bibfield  {author} {\bibinfo {author} {\bibfnamefont {J.~M.}\ \bibnamefont
  {Yao}},\ }\bibinfo {title} {Symmetry restoration methods},\ in\ \href
  {https://doi.org/10.1007/978-981-15-8818-1_18-1} {\emph {\bibinfo {booktitle}
  {Handbook of Nuclear Physics}}},\ \bibinfo {editor} {edited by\ \bibinfo
  {editor} {\bibfnamefont {I.}~\bibnamefont {Tanihata}}, \bibinfo {editor}
  {\bibfnamefont {H.}~\bibnamefont {Toki}},\ and\ \bibinfo {editor}
  {\bibfnamefont {T.}~\bibnamefont {Kajino}}}\ (\bibinfo  {publisher} {Springer
  Nature Singapore},\ \bibinfo {address} {Singapore},\ \bibinfo {year} {2020})\
  pp.\ \bibinfo {pages} {1--36}\BibitemShut {NoStop}%
\bibitem [{\citenamefont {Yao}\ \emph {et~al.}(2022)\citenamefont {Yao},
  \citenamefont {Meng}, \citenamefont {Niu},\ and\ \citenamefont
  {Ring}}]{Yao22}%
  \BibitemOpen
  \bibfield  {author} {\bibinfo {author} {\bibfnamefont {J.~M.}\ \bibnamefont
  {Yao}}, \bibinfo {author} {\bibfnamefont {J.}~\bibnamefont {Meng}}, \bibinfo
  {author} {\bibfnamefont {Y.~F.}\ \bibnamefont {Niu}},\ and\ \bibinfo {author}
  {\bibfnamefont {P.}~\bibnamefont {Ring}},\ }\bibfield  {title} {\bibinfo
  {title} {Beyond-mean-field approaches for nuclear neutrinoless double beta
  decay in the standard mechanism},\ }\href
  {https://doi.org/10.1016/j.ppnp.2022.103965} {\bibfield  {journal} {\bibinfo
  {journal} {Prog. Part. Nucl. Phys.}\ }\textbf {\bibinfo {volume} {126}},\
  \bibinfo {pages} {103965} (\bibinfo {year} {2022})}\BibitemShut {NoStop}%
\bibitem [{\citenamefont {Gambhir}\ \emph {et~al.}(1990)\citenamefont
  {Gambhir}, \citenamefont {Ring},\ and\ \citenamefont {Thimet}}]{Gambhir90}%
  \BibitemOpen
  \bibfield  {author} {\bibinfo {author} {\bibfnamefont {Y.~K.}\ \bibnamefont
  {Gambhir}}, \bibinfo {author} {\bibfnamefont {P.}~\bibnamefont {Ring}},\ and\
  \bibinfo {author} {\bibfnamefont {A.}~\bibnamefont {Thimet}},\ }\bibfield
  {title} {\bibinfo {title} {Relativistic mean field theory for finite
  nuclei},\ }\href {https://doi.org/10.1016/0003-4916(90)90330-Q} {\bibfield
  {journal} {\bibinfo  {journal} {Ann. Phys. (NY)}\ }\textbf {\bibinfo {volume}
  {198}},\ \bibinfo {pages} {132} (\bibinfo {year} {1990})}\BibitemShut
  {NoStop}%
\bibitem [{\citenamefont {Ring}(1996)}]{Ring96}%
  \BibitemOpen
  \bibfield  {author} {\bibinfo {author} {\bibfnamefont {P.}~\bibnamefont
  {Ring}},\ }\bibfield  {title} {\bibinfo {title} {Relativistic mean field
  theory in finite nuclei},\ }\href
  {https://doi.org/10.1016/0146-6410(96)00054-3} {\bibfield  {journal}
  {\bibinfo  {journal} {Prog. Part. Nucl. Phys.}\ }\textbf {\bibinfo {volume}
  {37}},\ \bibinfo {pages} {193} (\bibinfo {year} {1996})}\BibitemShut
  {NoStop}%
\bibitem [{\citenamefont {Meng}\ \emph {et~al.}(2006)\citenamefont {Meng},
  \citenamefont {Toki}, \citenamefont {Zhou} \emph {et~al.}}]{Meng06}%
  \BibitemOpen
  \bibfield  {author} {\bibinfo {author} {\bibfnamefont {J.}~\bibnamefont
  {Meng}}, \bibinfo {author} {\bibfnamefont {H.}~\bibnamefont {Toki}}, \bibinfo
  {author} {\bibfnamefont {S.~G.}\ \bibnamefont {Zhou}}, \emph {et~al.},\
  }\bibfield  {title} {\bibinfo {title} {Relativistic continuum {Hartree}
  {Bogoliubov} theory for ground-state properties of exotic nuclei},\ }\href
  {https://doi.org/10.1016/j.ppnp.2005.06.001} {\bibfield  {journal} {\bibinfo
  {journal} {Prog. Part. Nucl. Phys.}\ }\textbf {\bibinfo {volume} {57}},\
  \bibinfo {pages} {470} (\bibinfo {year} {2006})}\BibitemShut {NoStop}%
\bibitem [{\citenamefont {Hill}\ and\ \citenamefont {Wheeler}(1953)}]{Hill53}%
  \BibitemOpen
  \bibfield  {author} {\bibinfo {author} {\bibfnamefont {D.~L.}\ \bibnamefont
  {Hill}}\ and\ \bibinfo {author} {\bibfnamefont {J.~A.}\ \bibnamefont
  {Wheeler}},\ }\bibfield  {title} {\bibinfo {title} {Nuclear constitution and
  the interpretation of fission phenomena},\ }\href
  {https://doi.org/10.1103/PhysRev.89.1102} {\bibfield  {journal} {\bibinfo
  {journal} {Phys. Rev.}\ }\textbf {\bibinfo {volume} {89}},\ \bibinfo {pages}
  {1102} (\bibinfo {year} {1953})}\BibitemShut {NoStop}%
\bibitem [{\citenamefont {Griffin}\ and\ \citenamefont
  {Wheeler}(1957)}]{Griffin57}%
  \BibitemOpen
  \bibfield  {author} {\bibinfo {author} {\bibfnamefont {J.~J.}\ \bibnamefont
  {Griffin}}\ and\ \bibinfo {author} {\bibfnamefont {J.~A.}\ \bibnamefont
  {Wheeler}},\ }\bibfield  {title} {\bibinfo {title} {Collective motions in
  nuclei by the method of generator coordinates},\ }\href
  {https://doi.org/10.1103/PhysRev.108.311} {\bibfield  {journal} {\bibinfo
  {journal} {Phys. Rev.}\ }\textbf {\bibinfo {volume} {108}},\ \bibinfo {pages}
  {311} (\bibinfo {year} {1957})}\BibitemShut {NoStop}%
\bibitem [{\citenamefont {Ring}\ and\ \citenamefont {Schuck}(1980)}]{Ring80}%
  \BibitemOpen
  \bibfield  {author} {\bibinfo {author} {\bibfnamefont {P.}~\bibnamefont
  {Ring}}\ and\ \bibinfo {author} {\bibfnamefont {P.}~\bibnamefont {Schuck}},\
  }\href@noop {} {\emph {\bibinfo {title} {The Nuclear Many-Body Problem}}}\
  (\bibinfo  {publisher} {Springer},\ \bibinfo {address} {Heidelberg},\
  \bibinfo {year} {1980})\BibitemShut {NoStop}%
\bibitem [{\citenamefont {{Yao}}\ \emph {et~al.}(2014)\citenamefont {{Yao}},
  \citenamefont {{Hagino}}, \citenamefont {{Li}}, \citenamefont {{Meng}},\ and\
  \citenamefont {{Ring}}}]{Yao14}%
  \BibitemOpen
  \bibfield  {author} {\bibinfo {author} {\bibfnamefont {J.~M.}\ \bibnamefont
  {{Yao}}}, \bibinfo {author} {\bibfnamefont {K.}~\bibnamefont {{Hagino}}},
  \bibinfo {author} {\bibfnamefont {Z.~P.}\ \bibnamefont {{Li}}}, \bibinfo
  {author} {\bibfnamefont {J.}~\bibnamefont {{Meng}}},\ and\ \bibinfo {author}
  {\bibfnamefont {P.}~\bibnamefont {{Ring}}},\ }\bibfield  {title} {\bibinfo
  {title} {{Microscopic benchmark study of triaxiality in low-lying states of
  $^{76}$Kr}},\ }\href {https://doi.org/10.1103/PhysRevC.89.054306} {\bibfield
  {journal} {\bibinfo  {journal} {Phys. Rev. C}\ }\textbf {\bibinfo {volume}
  {89}},\ \bibinfo {eid} {054306} (\bibinfo {year} {2014})}\BibitemShut
  {NoStop}%
\bibitem [{\citenamefont {Zhao}\ \emph {et~al.}(2010)\citenamefont {Zhao},
  \citenamefont {Li}, \citenamefont {Yao},\ and\ \citenamefont
  {Meng}}]{Zhao10}%
  \BibitemOpen
  \bibfield  {author} {\bibinfo {author} {\bibfnamefont {P.~W.}\ \bibnamefont
  {Zhao}}, \bibinfo {author} {\bibfnamefont {Z.~P.}\ \bibnamefont {Li}},
  \bibinfo {author} {\bibfnamefont {J.~M.}\ \bibnamefont {Yao}},\ and\ \bibinfo
  {author} {\bibfnamefont {J.}~\bibnamefont {Meng}},\ }\bibfield  {title}
  {\bibinfo {title} {New parametrization for the nuclear covariant energy
  density functional with a point-coupling interaction},\ }\href
  {https://doi.org/10.1103/PhysRevC.82.054319} {\bibfield  {journal} {\bibinfo
  {journal} {Phys. Rev. C}\ }\textbf {\bibinfo {volume} {82}},\ \bibinfo
  {pages} {054319} (\bibinfo {year} {2010})}\BibitemShut {NoStop}%
\bibitem [{\citenamefont {Zhang}\ \emph {et~al.}(2022)\citenamefont {Zhang},
  \citenamefont {Cheoun}, \citenamefont {Choi} \emph {et~al.}}]{Zhang22}%
  \BibitemOpen
  \bibfield  {author} {\bibinfo {author} {\bibfnamefont {K.}~\bibnamefont
  {Zhang}}, \bibinfo {author} {\bibfnamefont {M.-K.}\ \bibnamefont {Cheoun}},
  \bibinfo {author} {\bibfnamefont {Y.-B.}\ \bibnamefont {Choi}}, \emph
  {et~al.} (\bibinfo {collaboration} {DRHBc Mass Table Collaboration}),\
  }\bibfield  {title} {\bibinfo {title} {{Nuclear mass table in deformed
  relativistic Hartree–Bogoliubov theory in continuum, I: Even-even
  nuclei}},\ }\href {https://doi.org/10.1016/j.adt.2022.101488} {\bibfield
  {journal} {\bibinfo  {journal} {At. Data Nucl. Data Tables}\ }\textbf
  {\bibinfo {volume} {144}},\ \bibinfo {pages} {101488} (\bibinfo {year}
  {2022})},\ \bibinfo {note} {\url{https://drhbctable.jcnp.org/}}\BibitemShut
  {NoStop}%
\bibitem [{\citenamefont {{Quan}}\ \emph {et~al.}(2017)\citenamefont {{Quan}},
  \citenamefont {{Chen}}, \citenamefont {{Li}}, \citenamefont
  {{Nik{\v{s}}i{\'c}}},\ and\ \citenamefont {{Vretenar}}}]{Quan17}%
  \BibitemOpen
  \bibfield  {author} {\bibinfo {author} {\bibfnamefont {S.}~\bibnamefont
  {{Quan}}}, \bibinfo {author} {\bibfnamefont {Q.}~\bibnamefont {{Chen}}},
  \bibinfo {author} {\bibfnamefont {Z.~P.}\ \bibnamefont {{Li}}}, \bibinfo
  {author} {\bibfnamefont {T.}~\bibnamefont {{Nik{\v{s}}i{\'c}}}},\ and\
  \bibinfo {author} {\bibfnamefont {D.}~\bibnamefont {{Vretenar}}},\ }\bibfield
   {title} {\bibinfo {title} {{Global analysis of quadrupole shape invariants
  based on covariant energy density functionals}},\ }\href
  {https://doi.org/10.1103/PhysRevC.95.054321} {\bibfield  {journal} {\bibinfo
  {journal} {Phys. Rev. C}\ }\textbf {\bibinfo {volume} {95}},\ \bibinfo {eid}
  {054321} (\bibinfo {year} {2017})}\BibitemShut {NoStop}%
\bibitem [{\citenamefont {Wu}\ \emph {et~al.}(2024)\citenamefont {Wu},
  \citenamefont {Cao}, \citenamefont {Zhao}, \citenamefont {Liu}, \citenamefont
  {Xiang},\ and\ \citenamefont {Zhou}}]{Wu24}%
  \BibitemOpen
  \bibfield  {author} {\bibinfo {author} {\bibfnamefont {X.-Y.}\ \bibnamefont
  {Wu}}, \bibinfo {author} {\bibfnamefont {J.-Z.}\ \bibnamefont {Cao}},
  \bibinfo {author} {\bibfnamefont {K.-N.}\ \bibnamefont {Zhao}}, \bibinfo
  {author} {\bibfnamefont {Z.-M.}\ \bibnamefont {Liu}}, \bibinfo {author}
  {\bibfnamefont {J.}~\bibnamefont {Xiang}},\ and\ \bibinfo {author}
  {\bibfnamefont {E.-F.}\ \bibnamefont {Zhou}},\ }\bibfield  {title} {\bibinfo
  {title} {Static or dynamic pear shapes in radioactive nucleus
  $^{224}\mathrm{Rn}$?},\ }\href {https://doi.org/10.1007/s41365-024-01578-z}
  {\bibfield  {journal} {\bibinfo  {journal} {Nucl. Sci. Tech.}\ }\textbf
  {\bibinfo {volume} {35}},\ \bibinfo {pages} {202} (\bibinfo {year}
  {2024})}\BibitemShut {NoStop}%
\bibitem [{\citenamefont {Krieger}\ \emph {et~al.}(1990)\citenamefont
  {Krieger}, \citenamefont {Bonche}, \citenamefont {Flocard}, \citenamefont
  {Quentin},\ and\ \citenamefont {Weiss}}]{Krieger90}%
  \BibitemOpen
  \bibfield  {author} {\bibinfo {author} {\bibfnamefont {S.~J.}\ \bibnamefont
  {Krieger}}, \bibinfo {author} {\bibfnamefont {P.}~\bibnamefont {Bonche}},
  \bibinfo {author} {\bibfnamefont {H.}~\bibnamefont {Flocard}}, \bibinfo
  {author} {\bibfnamefont {P.}~\bibnamefont {Quentin}},\ and\ \bibinfo {author}
  {\bibfnamefont {M.~S.}\ \bibnamefont {Weiss}},\ }\bibfield  {title} {\bibinfo
  {title} {An improved pairing interaction for mean field calculations using
  skyrme potentials*},\ }\href {https://doi.org/10.1016/0375-9474(90)90035-K}
  {\bibfield  {journal} {\bibinfo  {journal} {Nucl. Phys. A}\ }\textbf
  {\bibinfo {volume} {517}},\ \bibinfo {pages} {275} (\bibinfo {year}
  {1990})}\BibitemShut {NoStop}%
\bibitem [{\citenamefont {Robledo}(2009)}]{Robledo09}%
  \BibitemOpen
  \bibfield  {author} {\bibinfo {author} {\bibfnamefont {L.~M.}\ \bibnamefont
  {Robledo}},\ }\bibfield  {title} {\bibinfo {title} {{Sign of the overlap of
  Hartree-Fock-Bogoliubov wave functions}},\ }\href
  {https://doi.org/10.1103/PhysRevC.79.021302} {\bibfield  {journal} {\bibinfo
  {journal} {Phys. Rev. C}\ }\textbf {\bibinfo {volume} {79}},\ \bibinfo
  {pages} {021302(R)} (\bibinfo {year} {2009})}\BibitemShut {NoStop}%
\bibitem [{\citenamefont {Shiga}\ \emph {et~al.}(2016)\citenamefont {Shiga},
  \citenamefont {Yoneda}, \citenamefont {Steppenbeck} \emph
  {et~al.}}]{Shiga16}%
  \BibitemOpen
  \bibfield  {author} {\bibinfo {author} {\bibfnamefont {Y.}~\bibnamefont
  {Shiga}}, \bibinfo {author} {\bibfnamefont {K.}~\bibnamefont {Yoneda}},
  \bibinfo {author} {\bibfnamefont {D.}~\bibnamefont {Steppenbeck}}, \emph
  {et~al.},\ }\bibfield  {title} {\bibinfo {title} {Investigating nuclear shell
  structure in the vicinity of $^{78}${Ni}: Low-lying excited states in the
  neutron-rich isotopes $^{82,82}${Zn}},\ }\href
  {https://doi.org/10.1103/PhysRevC.93.024320} {\bibfield  {journal} {\bibinfo
  {journal} {Phys. Rev. C}\ }\textbf {\bibinfo {volume} {93}},\ \bibinfo
  {pages} {024320} (\bibinfo {year} {2016})}\BibitemShut {NoStop}%
\bibitem [{\citenamefont {Wang}\ \emph {et~al.}(2021)\citenamefont {Wang},
  \citenamefont {Huang}, \citenamefont {Kondev}, \citenamefont {Audi},\ and\
  \citenamefont {Naimi}}]{wang2021ame}%
  \BibitemOpen
  \bibfield  {author} {\bibinfo {author} {\bibfnamefont {M.}~\bibnamefont
  {Wang}}, \bibinfo {author} {\bibfnamefont {W.~J.}\ \bibnamefont {Huang}},
  \bibinfo {author} {\bibfnamefont {F.~G.}\ \bibnamefont {Kondev}}, \bibinfo
  {author} {\bibfnamefont {G.}~\bibnamefont {Audi}},\ and\ \bibinfo {author}
  {\bibfnamefont {S.}~\bibnamefont {Naimi}},\ }\bibfield  {title} {\bibinfo
  {title} {The ame 2020 atomic mass evaluation (ii). tables, graphs and
  references},\ }\href {https://doi.org/10.1088/1674-1137/abddaf} {\bibfield
  {journal} {\bibinfo  {journal} {Chin. Phys. C}\ }\textbf {\bibinfo {volume}
  {45}},\ \bibinfo {pages} {030003} (\bibinfo {year} {2021})}\BibitemShut
  {NoStop}%
\bibitem [{\citenamefont {{National Nuclear Data Center}}()}]{NNDC3}%
  \BibitemOpen
  \bibfield  {author} {\bibinfo {author} {\bibnamefont {{National Nuclear Data
  Center}}},\ }\href@noop {} {\bibinfo {title} {{Brookhaven National
  Laboratory}}},\ \bibinfo {howpublished}
  {\url{https://www.nndc.bnl.gov/nudat3/}}\BibitemShut {NoStop}%
\bibitem [{\citenamefont {Meng}\ \emph {et~al.}(2002)\citenamefont {Meng},
  \citenamefont {Toki}, \citenamefont {Zeng} \emph {et~al.}}]{Meng02}%
  \BibitemOpen
  \bibfield  {author} {\bibinfo {author} {\bibfnamefont {J.}~\bibnamefont
  {Meng}}, \bibinfo {author} {\bibfnamefont {H.}~\bibnamefont {Toki}}, \bibinfo
  {author} {\bibfnamefont {J.~Y.}\ \bibnamefont {Zeng}}, \emph {et~al.},\
  }\bibfield  {title} {\bibinfo {title} {{Giant halo at the neutron drip line
  in Ca isotopes in relativistic continuum Hartree-Bogoliubov theory}},\ }\href
  {https://doi.org/10.1103/PhysRevC.65.041302} {\bibfield  {journal} {\bibinfo
  {journal} {Phys. Rev. C}\ }\textbf {\bibinfo {volume} {65}},\ \bibinfo
  {pages} {041302(R)} (\bibinfo {year} {2002})}\BibitemShut {NoStop}%
\bibitem [{\citenamefont {{Neufcourt}}\ \emph {et~al.}(2019)\citenamefont
  {{Neufcourt}}, \citenamefont {{Cao}}, \citenamefont {{Nazarewicz}},
  \citenamefont {{Olsen}},\ and\ \citenamefont {{Viens}}}]{Neufcourt19}%
  \BibitemOpen
  \bibfield  {author} {\bibinfo {author} {\bibfnamefont {L.}~\bibnamefont
  {{Neufcourt}}}, \bibinfo {author} {\bibfnamefont {Y.}~\bibnamefont {{Cao}}},
  \bibinfo {author} {\bibfnamefont {W.}~\bibnamefont {{Nazarewicz}}}, \bibinfo
  {author} {\bibfnamefont {E.}~\bibnamefont {{Olsen}}},\ and\ \bibinfo {author}
  {\bibfnamefont {F.}~\bibnamefont {{Viens}}},\ }\bibfield  {title} {\bibinfo
  {title} {{Neutron Drip Line in the Ca Region from Bayesian Model
  Averaging}},\ }\href {https://doi.org/10.1103/PhysRevLett.122.062502}
  {\bibfield  {journal} {\bibinfo  {journal} {Phys. Rev. Lett.}\ }\textbf
  {\bibinfo {volume} {122}},\ \bibinfo {pages} {062502} (\bibinfo {year}
  {2019})}\BibitemShut {NoStop}%
\bibitem [{\citenamefont {Li}\ \emph {et~al.}(2020)\citenamefont {Li},
  \citenamefont {Hu}, \citenamefont {Wu} \emph {et~al.}}]{Li20}%
  \BibitemOpen
  \bibfield  {author} {\bibinfo {author} {\bibfnamefont {J.~G.}\ \bibnamefont
  {Li}}, \bibinfo {author} {\bibfnamefont {B.~S.}\ \bibnamefont {Hu}}, \bibinfo
  {author} {\bibfnamefont {Q.}~\bibnamefont {Wu}}, \emph {et~al.},\ }\bibfield
  {title} {\bibinfo {title} {Neutron-rich calcium isotopes within realistic
  gamow shell model calculations with continuum coupling},\ }\href
  {https://doi.org/10.1103/PhysRevC.102.034302} {\bibfield  {journal} {\bibinfo
   {journal} {Phys. Rev. C}\ }\textbf {\bibinfo {volume} {102}},\ \bibinfo
  {pages} {034302} (\bibinfo {year} {2020})}\BibitemShut {NoStop}%
\bibitem [{\citenamefont {Cortés}\ \emph {et~al.}(2018)\citenamefont
  {Cortés}, \citenamefont {Doornenbal}, \citenamefont {Dupuis} \emph
  {et~al.}}]{Cortes18}%
  \BibitemOpen
  \bibfield  {author} {\bibinfo {author} {\bibfnamefont {M.~L.}\ \bibnamefont
  {Cortés}}, \bibinfo {author} {\bibfnamefont {P.}~\bibnamefont {Doornenbal}},
  \bibinfo {author} {\bibfnamefont {M.}~\bibnamefont {Dupuis}}, \emph
  {et~al.},\ }\bibfield  {title} {\bibinfo {title} {{Inelastic scattering of
  neutron-rich Ni and Zn isotopes off a proton target}},\ }\href
  {https://doi.org/10.1103/PhysRevC.97.044315} {\bibfield  {journal} {\bibinfo
  {journal} {Phys. Rev. C}\ }\textbf {\bibinfo {volume} {97}},\ \bibinfo
  {pages} {044315} (\bibinfo {year} {2018})}\BibitemShut {NoStop}%
\bibitem [{\citenamefont {Yurkewicz}\ \emph {et~al.}(2004)\citenamefont
  {Yurkewicz}, \citenamefont {Bazin}, \citenamefont {Brown} \emph
  {et~al.}}]{Yurkewicz04}%
  \BibitemOpen
  \bibfield  {author} {\bibinfo {author} {\bibfnamefont {K.~L.}\ \bibnamefont
  {Yurkewicz}}, \bibinfo {author} {\bibfnamefont {D.}~\bibnamefont {Bazin}},
  \bibinfo {author} {\bibfnamefont {B.~A.}\ \bibnamefont {Brown}}, \emph
  {et~al.},\ }\bibfield  {title} {\bibinfo {title} {Nuclear structure in the
  vicinity of {$N=Z=28$} $^{56}\mathrm{Ni}$},\ }\href
  {https://doi.org/10.1103/PhysRevC.70.054319} {\bibfield  {journal} {\bibinfo
  {journal} {Phys. Rev. C}\ }\textbf {\bibinfo {volume} {70}},\ \bibinfo
  {pages} {054319} (\bibinfo {year} {2004})}\BibitemShut {NoStop}%
\bibitem [{\citenamefont {Sorlin}\ \emph {et~al.}(2002)\citenamefont {Sorlin},
  \citenamefont {Leenhardt}, \citenamefont {Donzaud} \emph
  {et~al.}}]{Sorlin02}%
  \BibitemOpen
  \bibfield  {author} {\bibinfo {author} {\bibfnamefont {O.}~\bibnamefont
  {Sorlin}}, \bibinfo {author} {\bibfnamefont {S.}~\bibnamefont {Leenhardt}},
  \bibinfo {author} {\bibfnamefont {C.}~\bibnamefont {Donzaud}}, \emph
  {et~al.},\ }\bibfield  {title} {\bibinfo {title} {$_{28}^{68}${Ni}$_{40}$:
  Magicity versus superfluidity},\ }\href
  {https://doi.org/10.1103/PhysRevLett.88.092501} {\bibfield  {journal}
  {\bibinfo  {journal} {Phys. Rev. Lett.}\ }\textbf {\bibinfo {volume} {88}},\
  \bibinfo {pages} {092501} (\bibinfo {year} {2002})}\BibitemShut {NoStop}%
\bibitem [{\citenamefont {Aoi}\ \emph {et~al.}(2009)\citenamefont {Aoi},
  \citenamefont {Takeshita}, \citenamefont {Suzuki} \emph {et~al.}}]{Aoi09}%
  \BibitemOpen
  \bibfield  {author} {\bibinfo {author} {\bibfnamefont {N.}~\bibnamefont
  {Aoi}}, \bibinfo {author} {\bibfnamefont {E.}~\bibnamefont {Takeshita}},
  \bibinfo {author} {\bibfnamefont {H.}~\bibnamefont {Suzuki}}, \emph
  {et~al.},\ }\bibfield  {title} {\bibinfo {title} {Development of large
  deformation in $^{62}\mathrm{Cr}$},\ }\href
  {https://doi.org/10.1103/PhysRevLett.102.012502} {\bibfield  {journal}
  {\bibinfo  {journal} {Phys. Rev. Lett.}\ }\textbf {\bibinfo {volume} {102}},\
  \bibinfo {pages} {012502} (\bibinfo {year} {2009})}\BibitemShut {NoStop}%
\bibitem [{\citenamefont {Rother}\ \emph {et~al.}(2011)\citenamefont {Rother},
  \citenamefont {Dewald}, \citenamefont {Iwasaki} \emph {et~al.}}]{Rother11}%
  \BibitemOpen
  \bibfield  {author} {\bibinfo {author} {\bibfnamefont {W.}~\bibnamefont
  {Rother}}, \bibinfo {author} {\bibfnamefont {A.}~\bibnamefont {Dewald}},
  \bibinfo {author} {\bibfnamefont {H.}~\bibnamefont {Iwasaki}}, \emph
  {et~al.},\ }\bibfield  {title} {\bibinfo {title} {Enhanced quadrupole
  collectivity at {$N=40$}: The case of neutron-rich fe isotopes},\ }\href
  {https://doi.org/10.1103/PhysRevLett.106.022502} {\bibfield  {journal}
  {\bibinfo  {journal} {Phys. Rev. Lett.}\ }\textbf {\bibinfo {volume} {106}},\
  \bibinfo {pages} {022502} (\bibinfo {year} {2011})}\BibitemShut {NoStop}%
\bibitem [{\citenamefont {Baugher}\ \emph {et~al.}(2012)\citenamefont
  {Baugher}, \citenamefont {Gade}, \citenamefont {Janssens} \emph
  {et~al.}}]{Baugher12}%
  \BibitemOpen
  \bibfield  {author} {\bibinfo {author} {\bibfnamefont {T.}~\bibnamefont
  {Baugher}}, \bibinfo {author} {\bibfnamefont {A.}~\bibnamefont {Gade}},
  \bibinfo {author} {\bibfnamefont {R.~V.~F.}\ \bibnamefont {Janssens}}, \emph
  {et~al.},\ }\bibfield  {title} {\bibinfo {title} {Intermediate-energy coulomb
  excitation of $^{58,60,62}${Cr}: The onset of collectivity toward {$N=40$}},\
  }\href {https://doi.org/10.1103/PhysRevC.86.011305} {\bibfield  {journal}
  {\bibinfo  {journal} {Phys. Rev. C}\ }\textbf {\bibinfo {volume} {86}},\
  \bibinfo {pages} {011305(R)} (\bibinfo {year} {2012})}\BibitemShut {NoStop}%
\end{thebibliography}
\end{document}